# Stability of stratified two-phase flows in inclined channels

I. Barmak[a)], A. Yu. Gelfgat, A. Ullmann, and N. Brauner

*School of Mechanical Engineering, Tel Aviv University, Tel Aviv 69978, Israel*

Linear stability of stratified gas-liquid and liquid-liquid plane-parallel flows in inclined channels is studied with respect to all wavenumber perturbations. The main objective is to predict parameter regions in which stable stratified configuration in inclined channels exists. Up to three distinct base states with different holdups exist in inclined flows, so that the stability analysis has to be carried out for each branch separately. Special attention is paid to the multiple solution regions to reveal the feasibility of non-unique stable stratified configurations in inclined channels. The stability boundaries of each branch of steady state solutions are presented on the flow pattern map and are accompanied by critical wavenumbers and spatial profiles of the most unstable perturbations. Instabilities of different nature are visualized by streamlines of the neutrally stable perturbed flows, consisting of the critical perturbation superimposed on the base flow. The present analysis confirms the existence of two stable stratified flow configurations in a region of low flow rates in countercurrent liquid-liquid flows. These configurations become unstable with respect to shear mode of instability. It was revealed that in slightly upward inclined flows the lower and middle solutions for the holdup are stable in a part of the triple solution region, while the upper solution is always unstable. In the case of downward flows, in the triple solution region, none of the solutions are stable with respect to short-wave perturbations. These flows are stable only in the single solution region at low flow rates of the heavy phase, where long-wave perturbations are the most unstable ones.

## I. INTRODUCTION

Stratified two-phase flows are widely encountered in various natural phenomena as well as in different fields of engineering, such as semiconductors, petroleum and plastics industries. Clearly, the stratified flow with a smooth interface is obtainable in an inclined channel only for certain ranges of operating parameters for which this regime is stable. Instability may result in the generation of waves at the interface (stratified-wavy regime) or may even lead to transition to other flow patterns (e.g., intermittent flow, core-annular flow). Therefore, stability analysis of smooth stratified flow should be studied to obtain the stability limits, to explore instability mechanisms, and to predict which flow pattern can be expected.

To avoid complexity of a rigorous analysis in the circular pipe geometry the stability boundaries in this kind of flows are commonly predicted based on the simplified one-dimensional Two-Fluid model (e. g., Barnea et al., 1980, Barnea, 1991, Brauner and Moalem Maron, 1992, Simmons and Hanratty, 2001, Ullmann and Brauner, 2006). Nevertheless, the predictions obtained via the Two-Fluid model are critically dependent on the closure relations used to account for the base flow and its interaction with the interfacial disturbances (e.g., steady and wave induced wall and interfacial shear stresses, velocity profile shape factors, see Kushnir et al., 2007, 2014). Furthermore, in the Two-Fluid model the stability analysis is limited only to long-wave perturbations.

---

[a)] Author to whom correspondence should be addressed. Electronic mail: ilyab@post.tau.ac.il

An alternative approach is to carry out a comprehensive stability analysis in the simpler two-plate geometry of two-layer plane Poiseuille flow, while considering all possible infinitesimal perturbations. This approach is helpful for better understanding of mechanisms involved in destabilization of stratified flows. Indeed, since the classical work of Yih (1967), where long-wave two-dimensional analysis was presented, the stability of stratified flow in the two-plate geometry has been extensively studied in the literature (e.g. Hooper and Boyd, 1983, Yiantsios and Higgins, 1988, Charru and Fabre, 1994, Ó Náraigh et al, 2014, Kaffel and Riaz, 2015). Most of the studies addressed horizontal flow driven by an imposed pressure gradient (see Barmak et al., 2016 and references therein), where the gravity-driven multiple solutions do not exist. Studies on inclined flows instabilities (Tilley et al., 1994; Boomkamp and Miesen, 1996; Amaouche et al., 2007; Vempati et al., 2010, Allouche et al., 2015) did not consider the issue of gravity-driven multiple solutions. However, there always exist two possible solutions for the holdup of steady countercurrent flow with a smooth interface, whereas in concurrent up- and down-flows three distinct steady state solutions can be obtained in a limited range of the flow parameters (e.g., Landman, 1991, Barnea and Taitel, 1992, Ullmann et al., 2003a, b, 2004). It is of the practical importance to introduce multiple solution regions on flow pattern and stability maps, since they are encountered for the operating conditions in many industrial processes. The existence of multiple holdups in liquid-liquid flows was verified experimentally (Ullmann et al., 2003a, b). Apparently, stability analysis is required to determine the range of parameters where the assumption of smooth interface is applicable to each of the solution branches. Such analysis may also reveal which holdups are feasible in multiple solution regions and the association of the stability boundaries with the flow pattern transition.

In this study, the linear stability of stratified two-phase flows in inclined channels with respect to arbitrary wavenumber disturbances is performed. Owing to a variety of governing parameters, the analysis is applied for predicting the stability boundaries on the flow pattern maps of several characteristic concurrent and countercurrent gas-liquid and liquid-liquid flows in channels with various inclinations, as well as taking into account all possible wavenumbers. To the best of our knowledge, such a comprehensive stability study has never been reported before.

## II. PROBLEM FORMULATION

The flow configuration of a stratified two-layer flow of two immiscible incompressible fluids in an inclined channel $(0 < \beta < \pi/2)$ is sketched in Figure 1. The flow, assumed isothermal and two-dimensional, is driven by an imposed pressure gradient and a component of the gravity in the $x$ direction. The interface between the fluids, labeled as $j = 1, 2$ (1 – lower phase, 2 – upper phase), is assumed to be flat in the undisturbed base flow state. Under this assumption, the position of the interface is obtained as additional unknown value of the steady state plane-parallel solution (see below).

The flow in each fluid is described by the continuity and momentum equations that are rendered dimensionless in the standard manner (see Kushnir et al., 2014), choosing for the scales of length and velocity the height of the upper layer $h_2$ and the interfacial velocity $u_i$, respectively. The time and the pressure are scaled by $h_2/u_i$, and $\rho_2 u_i^2$, respectively. The dimensionless continuity and momentum equations are

$$div\,\mathbf{u}_j = 0,$$

$$\frac{\partial \mathbf{u}_j}{\partial t} + (\mathbf{u}_j \cdot \nabla)\mathbf{u}_j = -\frac{\rho_1}{r\rho_j}\nabla p_j + \frac{1}{\operatorname{Re}_2}\frac{\rho_1}{r\rho_j}\frac{m\mu_j}{\mu_1}\Delta \mathbf{u}_j + \frac{\sin\beta}{\operatorname{Fr}_2}\mathbf{e}_x - \frac{\cos\beta}{\operatorname{Fr}_2}\mathbf{e}_y, \qquad (1)$$

where $\mathbf{u}_j = (u_j, v_j)$ and $p_j$ are the velocity and pressure of the fluid $j$, $\rho_j$ and $\mu_j$ are the corresponding density and dynamic viscosity; $\mathbf{e}_x$ and $\mathbf{e}_y$ are the unit vectors in the direction of the x- and y-axes.

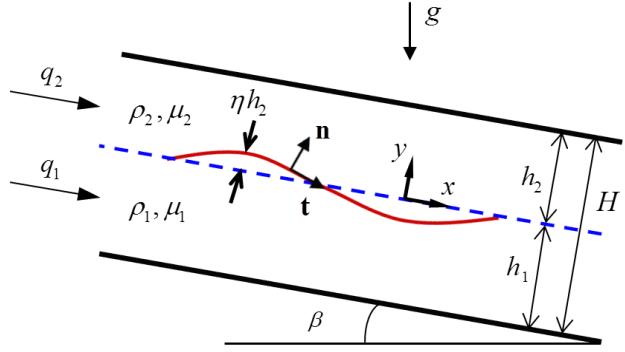

FIG. 1. Configuration of stratified two-layer flow in an inclined channel.

In the dimensionless formulation the lower and upper phases occupy the regions $-n \leq y \leq 0$, and $0 \leq y \leq 1$, respectively, $n = h_1/h_2$. Other dimensionless parameters: $\operatorname{Re}_{1,2} = \rho_{1,2} u_i h_2/\mu_{1,2}$ ($\operatorname{Re}_1 = \operatorname{Re}_2 r/m$) are the Reynolds numbers of each of the phases, $\operatorname{Fr}_2 = u_i^2/gh_2$ is the Froude number, $r = \rho_1/\rho_2$ and $m = \mu_1/\mu_2$ are the density and viscosity ratios. Gravity is acting in the downward direction $(-\mathbf{e}_y \cos\beta)$ and therefore has components in the streamwise and transverse directions, seen in the appearance of two terms with Froude number in Eq. (1).

The velocities satisfy the no-slip boundary conditions at the channel walls

$$\mathbf{u}_1(y = -n) = 0, \quad \mathbf{u}_2(y = 1) = 0. \qquad (2)$$

Boundary conditions at the interface $y = \eta(x, t)$ require continuity of velocity components and the tangential stresses, and jump of the normal stress due to the surface tension (the square brackets denote the jump of expression value across the interface)

$$\mathbf{u}_1(y = 0) = \mathbf{u}_2(y = 0), \qquad (3)$$

$$[\mathbf{t} \cdot \mathbf{T} \cdot \mathbf{n}] = \left[\frac{m\mu}{\mu_1}\left\{\left(\frac{\partial u}{\partial y} + \frac{\partial v}{\partial x}\right)\left(1 - \left(\frac{\partial \eta}{\partial x}\right)^2\right) - 4\frac{\partial u}{\partial x}\frac{\partial \eta}{\partial x}\right\}\right] = 0, \qquad (4)$$

$$[\mathbf{n} \cdot \mathbf{T} \cdot \mathbf{n}] = \left[ p + \frac{m\mu}{\mu_1} \frac{2\text{Re}_2^{-1}}{1 + \left(\frac{\partial \eta}{\partial x}\right)^2} \left( \frac{\partial u}{\partial x} \left(1 - \left(\frac{\partial \eta}{\partial x}\right)^2\right) + \left(\frac{\partial u}{\partial y} + \frac{\partial v}{\partial x}\right) \frac{\partial \eta}{\partial x} \right) \right]$$

$$= \text{We}_2^{-1} \frac{\frac{\partial^2 \eta}{\partial x^2}}{\left(1 + \left(\frac{\partial \eta}{\partial x}\right)^2\right)^{3/2}},$$

(5)

where **n** is the unit normal vector pointing from lower into upper phase, **t** is the unit vector tangent to the interface, **T** is the stress tensor. An additional dimensionless parameter $We_2 = \rho_2 h_2 u_i^2 / \sigma$ is the Weber number, and $\sigma$ is the surface tension coefficient.

Additionally, the interface displacement and the normal velocity components at the interface satisfy the kinematic boundary condition

$$v_j = \frac{D\eta}{Dt} = \frac{\partial \eta}{\partial t} + u_j \frac{\partial \eta}{\partial x}.$$

(6)

## III. THE BASE FLOW

### A. Steady velocity profile and pressure gradient

The base flow is assumed to be steady, laminar, and fully developed. Assuming that the velocity $U(y)$ is parallel to the channel walls and varies only with the cross-section coordinate $y$, the exact steady state solution can be found (see e.g., Ullmann et al., 2003). The solution yields the steady (dimensionless) velocity profiles

$$U_1 = 1 + a_1 y + b_1 y^2 \text{ for } -n \le y \le 0, \quad U_2 = 1 + a_2 y + b_2 y^2 \text{ for } 0 \le y \le 1,$$

(7)

where

$$a_1 = \frac{a_2}{m}, \quad a_2 = \frac{m - n^2 + n\tilde{Y}}{n^2 + n}, \quad b_1 = -\frac{m + n - \tilde{Y}}{(n^2 + n)m}, \quad b_2 = -\frac{m + n + n\tilde{Y}}{n^2 + n},$$

$$\tilde{Y} = \frac{n(1-r)\text{Re}_2 \sin \beta}{2 \text{Fr}_2} = \frac{Y(h-1)^2 (4h + m - 2mh + mh^2 - h^2)}{(Yh^4 - 3Yh^3 + 3Yh^2 - Yh - 1)}.$$

Note that, given three dimensionless parameters $n$, $m$, $\tilde{Y}$, the base flow characteristics can be determined. However, since these parameters are based on the unknown lower (heavy) phase holdup, $h = h_1 / H$, and the interfacial velocity, it is convenient to use the other common parameters for two-phase flow, which are based on the specified operational conditions. The base flow solution is fully determined by three following parameters: the Martinelli parameter $X^2 = (-dP/dx)_{1S} / (-dP/dx)_{2S} = m \cdot q$, the flow rate ratio $q = q_1 / q_2$, and also the inclination parameter $Y = \rho_2 (r-1) g \sin \beta / (-dP/dx)_{2S}$. Here $q_j$ is the feed flow rate of phase $j$ and $(-dP/dx)_{jS} = 12 \mu_j q_j / H^3$ is the corresponding superficial pressure drop for single phase flow channel, where $H = h_1 + h_2$.

To apply collocation method based on the Chebyshev polynomials (defined in the interval $[0,1]$), a new coordinate $y_1 = (y+n)/n$ $(0 \leq y_1 \leq 1)$ should be introduced for the part of the channel occupied by the lower phase, while $y_2 = y$ $(0 \leq y_2 \leq 1)$ for the upper phase remains unchanged. After substitution of the new coordinate the velocity profile in the lower phase reads

$$U_1 = 1 + a_1 y + b_1 y^2 = 1 + a_1 (y_1 - 1) n + b_1 (y_1 - 1)^2 n^2$$
$$= \underbrace{1 - a_1 n + b_1 n^2}_{\tilde{c}_1} + \underbrace{(a_1 n - 2 b_1 n^2)}_{\tilde{a}_1} y_1 + \underbrace{b_1 n^2}_{\tilde{b}_1} y_1^2. \qquad (8)$$

The holdup can be found by solving the following algebraic equation $F(Y, q, m, h) = 0$ (e.g., Ullmann et al., 2003a):

$$Y - \frac{mq(1-h)^2 \left[(1+2h)m + (1-m)h(4-h)\right] - h^2 \left[(3-2h)m + (1-m)h^2\right]}{4h^3 (1-h)^3 \left[h + m(1-h)\right]} = 0, \qquad (9)$$

or, alternatively, as the 7-th order algebraic equation in $h$

$$4Y(1-m)h^7 + 4Y(4m-3)h^6 + 12Y(1-2m)h^5$$
$$+ \left[(X^2 + 1) \cdot (m-1) + 4Y(4m-1)\right] h^4 + \left[2X^2 (3-2m) + 2m(1-2Y)\right] h^3 \qquad (10)$$
$$+ \left[3X^2 (2m-3) - 3m\right] h^2 + 4X^2 (1-m) h + X^2 m = 0.$$

The dimensionless pressure drop can be calculated as

$$\tilde{P} = \frac{dP/dx - \rho_2 g \sin \beta}{(-dP/dx)_{2S}} = \frac{3mq(1-h)^2 - 4mh(1-h) - h^2}{4\tilde{h}(1-h)^2 \left[(1+2h)m + (1-m)h(4-h) - 3h\right]}. \qquad (11)$$

The interfacial velocity can also be found

$$\tilde{u}_i = \frac{u_i}{U_{2S}} = \frac{6h(1-h)(Yh - \tilde{P})}{m(1-h) + h}, \qquad (12)$$

where $U_{2S} = q_2 / H$ is the superficial velocity of the upper phase.

Henceforth, we use the superficial velocity and the flow rate concepts interchangeable due to consideration of channels of constant height.

**B. Multiple holdup solution region**

Equation (10) yields a single solution for the holdup for horizontal flows $(Y=0)$, and its stability was the subject of our previous paper (Barmak et al., 2016). On the other hand, the important feature of flows in inclined channels, which are studied in the present work, is the existence of several stratified flow configurations (with different holdups) for fixed operational conditions that result from the solution of Equation (10). Two different holdups are obtained in the whole range of existence of flow with countercurrent feed (the heavy phase flows downward, the light one flows upward; $X^2<0, Y<0$). In concurrent flow $(X^2>0)$ a triple solution is predicted only in a limited range of flow parameters, typically at relatively high $X^2$ $(q_2 \ll q_1)$ for concurrent down-flow $(X^2>0, Y>0)$ and for small positive values of $X^2$ $(q_2 \gg q_1)$ for upward concurrent flow (where $X^2>0, Y<0$). Otherwise, there is a single solution for concurrent inclined flows. The feasibility of obtaining multiple holdups in inclined channels was validated experimentally in the works of Ullmann et al. (2003a, b, 2004). Nonuniqueness of the (smooth) stratified flow configuration, however, is feasible only if at least two distinct solutions are stable. Computation of the stability limits of the solution branches is the main objective of the present study.

In this paper stability analysis is carried out for the whole range of operation conditions. At the preliminary stage, the boundaries of multiple solution regions and possible base flow solutions should be found from Eq. (10). For a particular fluids (known viscosity ratio $m$) the range of the Martinelli parameters $X^2$ corresponding to multiple solutions is bounded by the values defined from $dX^2/dh=0$ for a specified value of the inclination factor $Y$. Using equation (9) for calculation of this derivative yields

$$Y = -\left[(m-1)^2 (h^4 - 2h^3) + 2m(m-1)h - m^2\right]$$
$$\times \left[(m-1)^2 h^4 - 2(2m^2 - 5m + 3)h^3 + 2(3m^2 - 6m + 2)h^2 + 4m(m-1)h + m^2\right]^{-1} \quad (13)$$
$$\times \left[2(h-1)^3 \cdot h\right]^{-1}.$$

The results obtained from Eq. (9) or (10) can be presented in the form of holdup dependence on $X^2$ (Figure 2) for a specified value of $Y$, whereas Eq. (13) can be used to show the boundaries of the multiple solution regions on a $Y-X^2$ diagram (Figure 3).

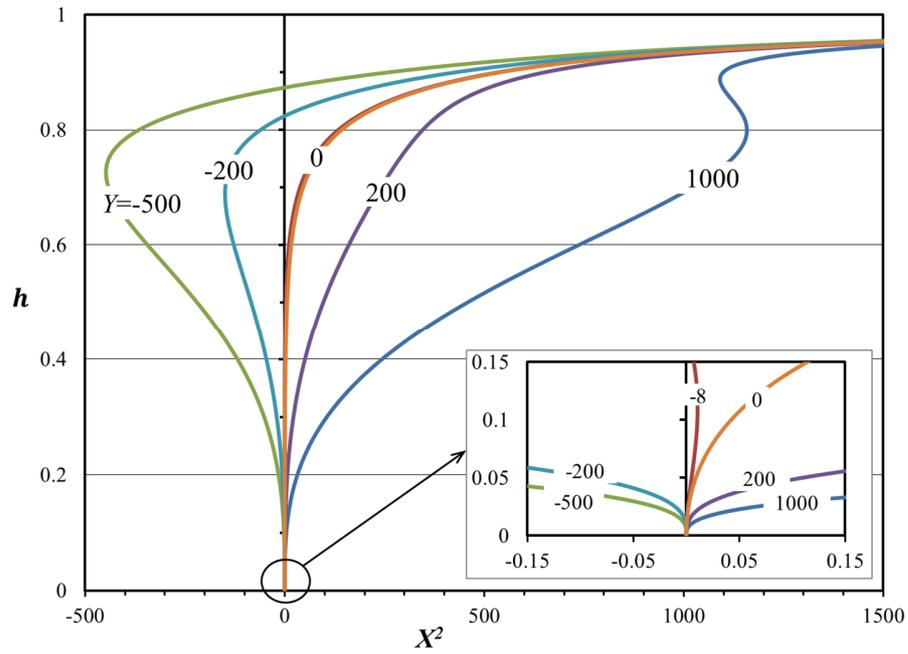

FIG. 2. Holdup solution for the base flow as a function of the Martinelli parameter $(m = 55)$.

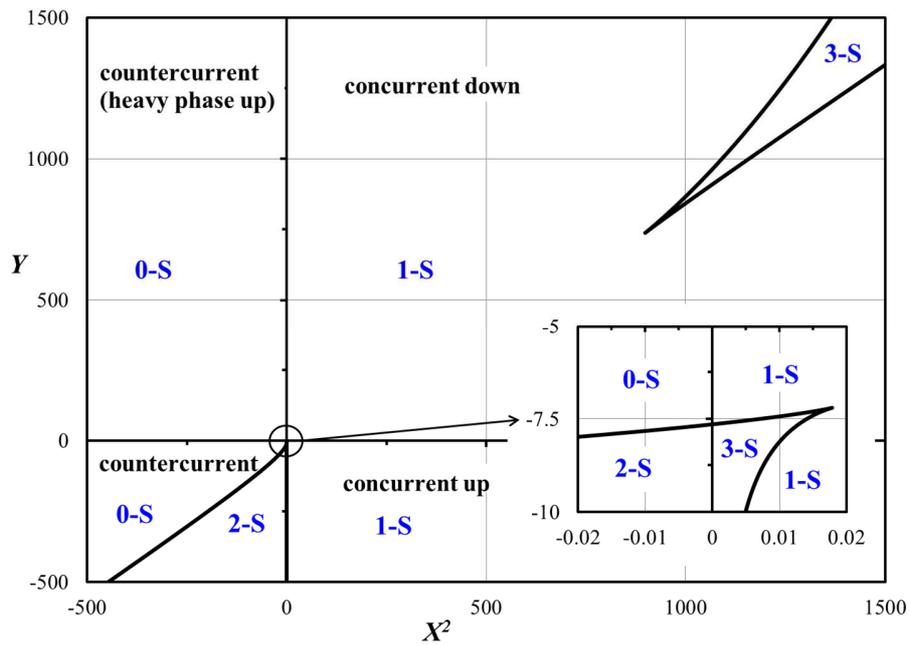

FIG. 3. Number of solutions diagram for concurrent and countercurrent flows $(m = 55)$; "0-S" denotes the region where no solution for the holdup exists, "1-S" – one solution for the holdup, etc.

**IV. LINEAR STABILITY**

In the following we study linear stability of the above plane-parallel solutions with respect to infinitesimal, two-dimensional disturbances. The perturbed velocities and pressure fields are written as $u_j = U_j + \tilde{u}_j$, $v_j = \tilde{v}_j$, $p_j = P_j + \tilde{p}_j$, and $\eta = \tilde{\eta}$ for the dimensionless disturbance of the interface. In this study only two-dimensional disturbances are considered, since they are shown to be the most unstable at least for horizontal stratified flows (e.g., Hesla et al., 1986). The disturbed velocities are conveniently represented by the corresponding stream function $\left( \tilde{u}_j = \partial \psi_j / \partial y ; \tilde{v}_j = -\partial \psi_j / \partial x \right)$, and an exponential dependence of the perturbation in time is assumed

$$\begin{pmatrix} \psi_j \\ \tilde{p}_j \\ \eta \end{pmatrix} = \begin{pmatrix} \phi_j(y) \\ f_j(y) \\ H_\eta \end{pmatrix} e^{(ikx+\lambda t)}, \quad \begin{pmatrix} \tilde{u}_j \\ \tilde{v}_j \end{pmatrix} = \begin{pmatrix} \phi'_j \\ -ik\phi_j \end{pmatrix} e^{(ikx+\lambda t)}, \tag{14}$$

where $\phi_j, f_j$ and $H_\eta$ are the perturbation amplitudes, $k$ is the dimensionless real wavenumber ($k = 2\pi h_2 / l_{wave}$, with $l_{wave}$ being the wavelength) and $\lambda$ is the complex time increment. After substitution and linearization of the original equations and boundary conditions (1)-(6), the problem is reduced to the Orr-Sommerfeld equations, written here in the eigenvalue problem form

$$0 \leq y_1 \leq 1 \; (-n \leq y \leq 0): \quad \lambda D_1 \phi_1 = \left[ ik\left( -U_1 D_1 + \frac{U_1''}{n^2} \right) + \frac{1}{\text{Re}_1} D_1^2 \right] \phi_1, \tag{15}$$

$$0 \leq y_2 \leq 1 \; (0 \leq y \leq 1): \quad \lambda D_2 \phi_2 = \left[ ik\left( -U_2 D_2 + U_2'' \right) + \frac{1}{\text{Re}_2} D_2^2 \right] \phi_2, \tag{16}$$

where

$$D_1 \phi = \frac{\phi''}{n^2} - k^2 \phi, \quad D_1^2 \phi = \frac{\phi^{IV}}{n^4} - 2k^2 \frac{\phi''}{n^2} + k^4 \phi,$$
$$D_2 \phi = \phi'' - k^2 \phi, \quad D_2^2 \phi = \phi^{IV} - 2k^2 \phi'' + k^4 \phi. \tag{17}$$

The linearized boundary conditions are obtained by means of Taylor expansions of $\eta$ around its unperturbed zero value (see Segal, 2008, for more details)

$$y_1 = 1, \; y_2 = 0: \quad \lambda H_\eta = -ik\left( \phi_2 + U_2 H_\eta \right), \tag{18}$$

$$\text{with} \quad H_\eta = \frac{\phi_2'(0) - \phi_1'(1)/n}{U_1'(1)/n - U_2'(0)},$$

$$y_1 = 1, y_2 = 0: \quad \begin{aligned} \lambda\left(r \cdot \frac{\phi_1'(1)}{n} - \phi_2'(0)\right) &= ik\left[-\left(k^2 \text{We}_2^{-1} + \frac{\cos\beta}{\text{Fr}_2}(r-1)\right) \cdot H_\eta \right. \\ &\quad + r\left(-U_1 \frac{\phi_1'(1)}{n} + \frac{U_1'(1)}{n}\phi_1(1)\right) + \left(U_2\phi_2'(0) - U_2'(0)\phi_2(0)\right)\Bigg] \\ &\quad + \frac{1}{\text{Re}_2}\left[m\left(\frac{\phi_1'''(1)}{n^3} - 3k^2 \frac{\phi_1'(1)}{n}\right) - \left(\phi_2'''(0) - 3k^2 \phi_2'(0)\right)\right], \end{aligned} \quad (19)$$

$$y_1 = 0\,(y = -n): \quad \phi_1 = \phi_1' = 0, \quad (20)$$

$$y_2 = 1: \quad \phi_2 = \phi_2' = 0, \quad (21)$$

$$y_1 = 1, y_2 = 0: \quad \phi_1(1) = \phi_2(0), \quad (22)$$

$$y_1 = 1, y_2 = 0: \quad m\left[\frac{\phi_1''(1)}{n^2} + k^2\phi_1(1) + \frac{U_1''}{n^2}H_\eta\right] = \phi_2''(0) + k^2\phi_2(0) + U_2''H_\eta. \quad (23)$$

In contrast to horizontal flow, in case of inclined flow the gradient of shear stress is not the same for the two layers, since the gravity has a streamwise component, and the displacement of the interface (the third terms on both sides) cannot be ignored in Eq. (23)

The temporal linear stability is studied by solving the differential system (15), (16) and (18)-(23) assuming an arbitrary wavenumber for each given set of the other parameters. Defining the time increment as a complex eigenvalue $\lambda = \lambda_R + i\lambda_I$, the growth rate of the perturbation is determined by $\lambda_R$. When $u_i > 0$, the flow is considered to be stable if the real parts of all eigenvalues are negative. On the other hand, the flow with $u_i < 0$ is stable only if all eigenvalues are positive, since the time is scaled by $h_2/u_i$, which becomes negative in this case. Neutral stability corresponds to $\max(\lambda_R) = 0$ for $u_i > 0$ ( $\min(\lambda_R) = 0$ for $u_i < 0$ ). The dimensionless phase speed of the perturbation is determined by a quantity $c_R = -\lambda_I / k$ (the dimensional one is $\hat{c}_R = c_R u_i$), where $\lambda_I$ is the wave angular frequency.

The stability problem is solved by applying the Chebyshev collocation method for discretization of the Orr-Sommerfeld equations and the boundary conditions (see details in Barmak et al., 2016) and by using the QR algorithm (Francis, 1962) for computation of the eigenvalues and eigenvectors. The numerical solution was verified by comparison with the solution for horizontal flow (presented in Barmak et al., 2016) for $\beta = 0°$ and with the long-wave asymptotic solution for inclined flow of Kushnir et al. (2014) (see Table I).

TABLE I. Comparison of the present numerical results and the asymptotic solution of Kushnir et al. (2014) for the critical superficial velocities. In both studies $U_{1S}$ was fixed, while $U_{2S}$ was varied.

|  | Inclined channel | $U_{1S}$, m/s | $U_{2S}$, m/s | |
|---|---|---|---|---|
|  |  |  | Asymptotic | Numerical |
| Liquid-liquid countercurrent flow | $H = 0.0144$ m $\beta = 10°$ $r = 1.033$ $m = 1.52$ | $6.944 \cdot 10^{-6}$ (lower solution) | -0.020054287 | -0.020051988 |
|  |  | 0.000208333 (lower) | -0.004775573 | -0.004774732 |
|  |  | 0.000277778 (lower) | -0.000098149 | -0.000101201 |
|  |  | $6.944 \cdot 10^{-6}$ (upper solution) | -0.000191477 | -0.000191556 |
|  |  | 0.000208333 (upper) | -0.000189808 | -0.000189999 |
|  |  | 0.019444444 (upper) | -0.000007620 | -0.000007629 |
| Air-water upward inclined flow | $H = 0.02$ m $\beta = 0.1°$ $r = 1000$ $m = 55$ | -0.000001 (single solution) | -0.004089976 | -0.004093802 |
|  |  | -0.001 (single solution) | -0.004071029 | -0.004075971 |
|  |  | -0.02 (single solution) | -0.003273662 | -0.003283526 |
|  |  | -0.000001 (lower solution) | -1.6244477 | -1.6241747 |
|  |  | -0.00005 (lower) | -3.4595159 | -3.4590146 |

## V. RESULTS AND DISCUSSION

Linear all wavelength stability analysis of liquid-liquid and gas-liquid flows in inclined channels is carried out to study stability limits of stratified flows and to reveal the destabilization mechanisms involved. In this geometry two more parameters $(Y, \beta)$ are needed in addition to the five dimensionless parameters (i.e., $m, q, r, \text{Re}_{2S}$ or $\text{Fr}_{2S}$, and $\text{We}_{2S}$), which govern the horizontal flow stability. Here $\text{Re}_{jS} = \rho_j U_{jS} H / \mu_j$ and $\text{Fr}_{jS} = U_{jS}^2 / (gH)$ are the superficial Reynolds and Froude numbers respectively, and $\text{We}_{jS} = \rho_j U_{jS}^2 H / \sigma$ is the superficial Weber number. In the present work the values of physical properties are set at several representative values, considering systems with various inclinations and different feeding configurations: concurrent upward and downward, and countercurrent flows. The stability of the flow is then studied by varying the flow rate of each of the phases. In all cases the upper layer is considered to be lighter than the lower layer, so that the Rayleigh-Taylor instability is not encountered. The main attention is focused on the operational conditions associated with multiple solutions for the holdup. Along with presenting accurate maps of stability, a comparison with the analytical results obtained for the long-wave limit is one of the tasks of this work. The most unstable perturbation for each particular configuration (described by the leading eigenfunction) is also reported, its pattern is discussed, which allows us to make some additional conclusions on the nature of instability.

There are several mechanisms, which can be responsible for destabilization of stratified flow with a smooth interface. Basically, they can be classified to the shear flow instability, which originates near the channel walls and is encountered also in single-phase Poiseuille flow (and associated with the onset of turbulence), and to the interfacial instability, which is associated with viscosity and/or density stratification (e.g., Yih, 1967; Kushnir et al., 2014). Viscosity stratification $(m \neq 1)$ is responsible for a discontinuity (jump) in the primary flow velocity gradient $U'_j$ across the interface, while density difference between the phases in inclined flows ($r > 1$) results in a jump in the curvature $U''_j$ of the base flow velocity profiles across the interface, such that $mU''_1 \neq U''_2$. In the general case of two fluids with different viscosities and densities, the effects of these jumps are coupled and lead to energy transfer from the base flow to the disturbed flow (Boomkamp and Miesen, 1996).

**A. Countercurrent flows**

As described above, two distinct stratified configurations with different holdups can be obtained in countercurrent flows (see Ullmann et al., 2003a). In countercurrent flow the heavy phase (lower layer) is flowing downward under the influence of gravity, while the light phase is moving upward due to a pressure gradient. The flow is opposed by interfacial friction between the phases, which depends on the relative countercurrent mean velocity of the phases $(U_1 - U_2)$. For a given fluids and inclination angle there is a maximum relative velocity that can be sustained in this type of flow. Beyond these conditions, the so-called "flooding phenomenon" (e.g., Bankoff and Lee, 1983) occurs (the minimal $X^2$ in Figure 2). Hence, there is an upper bound (flooding curve) for countercurrent on the flow pattern map (e.g., see Figure 4).

The non-uniqueness of solution encountered in countercurrent flow should be approached with extra care to ensure that the assumed configurations of smooth stratified flows are stable and indeed correspond to realizable physical configurations. The properties of the liquid-liquid system studied in Figures 4-10 are the same as those used in the experimental study of Ullmann et al. (2003a) ($r = 1.033$, $m = 1.52$, $\rho_2 = 916.6 \, \text{kg/m}^3$, $\mu_2 = 0.0024 \, \text{Pa} \cdot \text{s}$, $\sigma = 0.03 \, \text{N/m}$, $H = 1.44 \, \text{cm}$). The stability boundaries for a 10° inclined countercurrent liquid-liquid flow, which result from the analysis that considers all wavenumber perturbations, are depicted in Figure 4. There are two neutral stability curves: one for the lower holdup solution, denoted as the Light-Phase Dominated (LPD) configuration, and the other for the upper holdup solution, denoted as the Heavy Phase Dominated (HPD) configuration. The analysis revealed that a long-wave perturbation $(k \to 0)$ is the critical one for the whole range of flow rates. Hence the stability boundaries coincide with those obtained by long-wave analysis (see Kushnir et al., 2014). Upon considering all wavenumber perturbations in the stability analysis, this result was found to be valid for all countercurrent liquid-liquid flows studied. Thus, the exact long-wave analytical solution can be conveniently utilized for identifying the stable regions in liquid-liquid countercurrent flows. It should be noted however, that for the LPD configuration a wide range of long and intermediate wavelength perturbations (normalized wavenumber $k_H = 2\pi H / l_{wave}$ is up to about 0.1) becomes unstable almost simultaneously upon crossing the stability boundary (e.g., Figure 5(a)). As shown in Figure 5(a), in the unstable region the most amplified wave (e.g., $k_H = 0.22$ for point $C_2$) does not necessarily correspond to the critical perturbation (e.g., $k \to 0$ for point C, which defines the stability boundary). On the other hand, for the HPD configuration (e.g., Figure 5(b)), only very long waves $(k \to 0)$ trigger flow instability. As demonstrated in Figure 5(b), a slight increase of the superficial velocity of one of the phases beyond the critical conditions (i.e., in the unstable region) results in positive values of the growth rate for a range of long-wave perturbations.

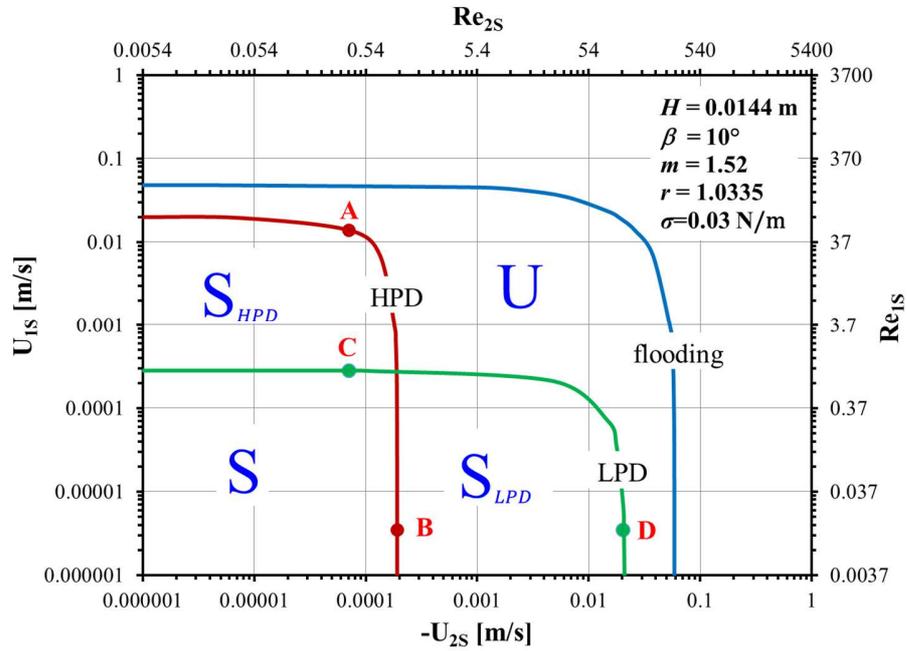

FIG. 4. Stability boundaries for countercurrent liquid-liquid flow $(\beta = 10°; m = 1.52; \rho_2 = 916.6 \, kg/m^3; \mu_2 = 2.4 \cdot 10^{-3} \, Pa \cdot s)$; HPD – heavy phase dominated solution (upper solution for the holdup); LPD – light phase dominated solution (lower solution for the holdup). Long-wave perturbation $(k \to 0)$ is the critical mode for the whole range of flow rates for both HPD and LPD.

In Figure 4, both solutions are observed to be stable for low flow rates. Figure 4 shows that the LPD stable region extends to higher light phase superficial velocities, whereas the HPD extends to higher heavy phase superficial velocities. At sufficiently high flow rates both solutions are unstable. This is further elucidated in Figure 6, where the stable solutions are indicated on the holdup curve $(h \text{ vs. } U_{2S})$ obtained for a constant heavy phase superficial velocity. It can be observed that stable countercurrent flows correspond either to very large holdup $(h > 0.9)$ for the HPD configuration, or to very small holdup $(h < 0.1)$ for the LPD configuration, hence in both cases the interface is located close to the channel wall. For the intermediate values of the holdup, a configuration with a smooth interface is not obtained, and a wavy interface is expected. The obtained theoretical results (in two-plate geometry) are in general agreement with the experimental observations of Zamir (2003) in a 10° inclined pipe flow, where the LPD and HPD modes were observed to be smooth stratified flow in the operational window of $U_{1S} = 0-0.005\,\text{m/s}$, $U_{2S} = 0-0.008\,\text{m/s}$. A further increase in the light phase flow rate results in the two unstable solutions, which eventually merge into a single unstable solution at the flooding point. It is worth noting, however, that in experiments conducted in a test section of a finite length the longest observable wave length is of the order of the channel length. Moreover, the residence time in the channel may be insufficient for the perturbation to grow to a detectable amplitude. Hence, the experimental stratified smooth flow region will extend to higher flow rates compared to theoretical predictions.

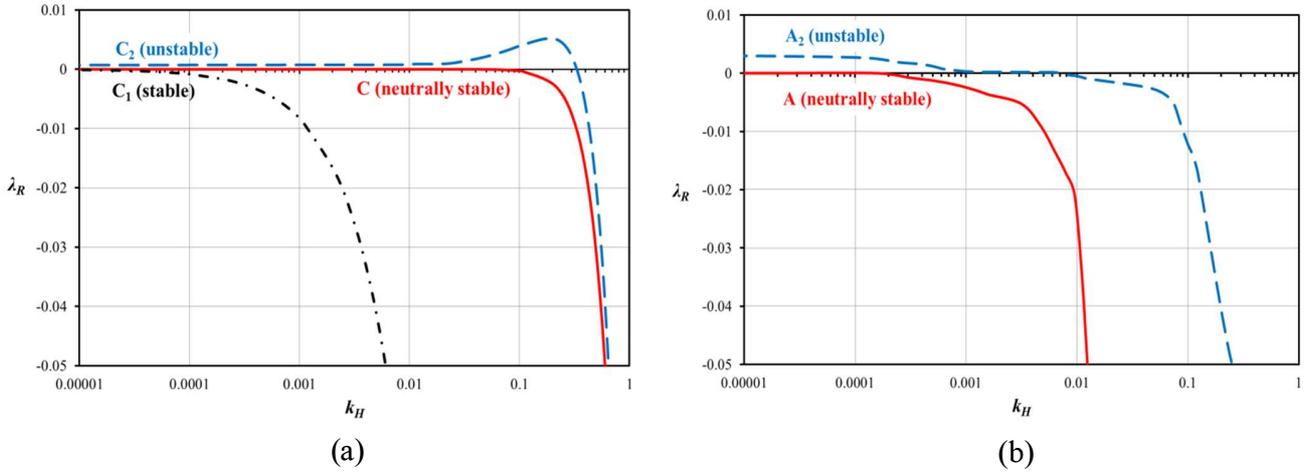

FIG. 5. Growth rate of perturbation vs. wavenumber. (a) For the LPD solution, at point C ($U_{1S} = 2.79 \cdot 10^{-4}\,\text{m/s}$; $U_{2S} = -7.08 \cdot 10^{-5}\,\text{m/s}$; $h = 0.069$, red solid line) on the neutral stability curve (see Figure 4) and in its vicinity in the stable region, at point $C_1$ ($U_{1S} = 2 \cdot 10^{-4}\,\text{m/s}$; $U_{2S} = -7.08 \cdot 10^{-5}\,\text{m/s}$; $h = 0.061$, black dash-dot line), and in the unstable region, at point $C_2$ ($U_{1S} = 3.5 \cdot 10^{-4}\,\text{m/s}$; $U_{2S} = -7.08 \cdot 10^{-5}\,\text{m/s}$; $h = 0.074$, blue dashed line). (b) For the HPD solution, at point A ($U_{1S} = 1.39 \cdot 10^{-2}\,\text{m/s}$; $U_{2S} = -7.08 \cdot 10^{-5}\,\text{m/s}$; $h = 0.937$, red solid line) on the neutral stability curve (see Figure 4) and in its vicinity in the unstable region, at point $A_2$ ($U_{1S} = 0.02\,\text{m/s}$; $U_{2S} = -7.08 \cdot 10^{-5}\,\text{m/s}$; $h = 0.919$, blue dashed line).

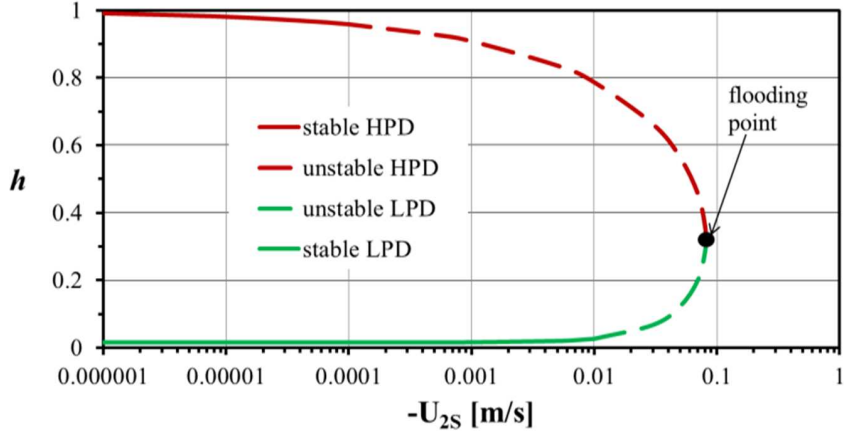

FIG. 6. The holdup vs. the light phase superficial velocity for constant heavy phase superficial velocity $\left(U_{1S}=10^{-5}\,\text{m/s}\right)$. Dashed lines represent unstable conditions.

    The patterns of critical perturbations of the stream function (eigenfunctions) for each solution give an additional physical insight on the location in the flow where the instability evolves. The absolute values of the eigenfunctions amplitude and their derivatives ($|\phi|$ and $|\phi'|$) are shown in Figures 7(a)-(h). For the purposes of further discussion, the base flow velocity profiles are also presented in the figures. Disturbances of the transverse and streamwise velocity components are proportional to the stream function disturbances and their derivative, respectively (see Eq. (14)). The value of eigenfunction derivative at zero y-coordinate is related to the interface displacement amplitude (see Eq. (18)). All the profiles are normalized by their values at the interface, except for $|\phi'|$, which is discontinuous across the interface and is normalized by $|\phi'_2(0)|$ in the light phase. The perturbation patterns are presented for several representative points (shown in Figure 4) for the HPD (points **A** and **B**) and the LPD (points **C** and **D**) solutions.

    The critical perturbation of the stream function has a maximum in the bulk of the heavy phase for the conditions corresponding to point **A** (see Figure 7(a)). At the same time the largest value of the eigenfunction derivative (the streamwise velocity perturbation) is observed at the interface on the lower layer side, while the secondary maximum (which is near the lower wall in single-phase Poiseuille flow) is shifted towards the interface (Figure 7(b)). Since the heavy phase flow rate is relatively high at point **A**, the interfacial velocity of the base flow is positive (i.e., directed downward), resulting in a small backflow region in the thin layer of the light phase in the vicinity of the interface. On the other hand, at point **B** (Figures 7(c) and (d)), where the flow rate of the heavy phase is relatively small, the heavy phase is dragged upward by the light phase (the interfacial velocity is negative), and a wide backflow region occupies the heavy phase domain near the interface. Nevertheless, the critical perturbation profile appears similar to that obtained at point **A**. It can be observed that the maximum of the transverse velocity perturbation ($|\phi|$) is shifted from the location of the maximal heavy phase velocity towards the interface and is near the location of zero base state velocity.

For the LPD case studies (points **C** and **D**), the maximum of the stream function perturbation is also observed in the bulk of the thicker layer (which is the light phase in these cases). At point **C** (Figures 7(e) and (f)) the light phase is dragged downward by the flow of the heavy (and more viscous) phase, and a backflow occupies a large part of the light phase domain. Similarly to case **B**, also in case **C**, the maximum of the transverse velocity corresponds to the location of zero base state velocity. Case **D**, which corresponds to relatively high flow rate of the light phase, is antisymmetrical to case **A**. In all four case studies, the maximum of the streamwise velocity perturbation is at the interface (in the dominating phase domain). The observed patterns unambiguously indicate a shear mode of instability. Instability is triggered by the shear in the dominating phase (i.e., the thicker layer), so that the most unstable regions are located near the wall and the interface.

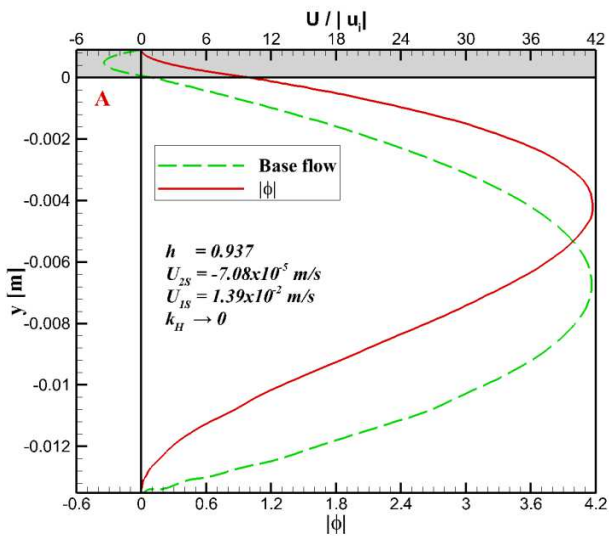
(a)

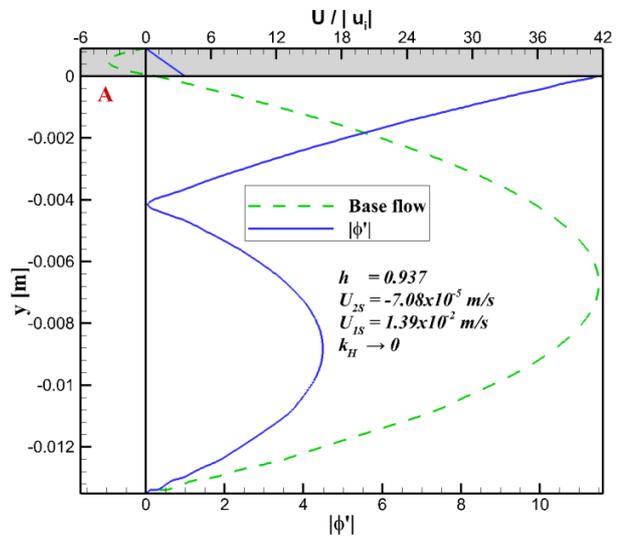
(b)

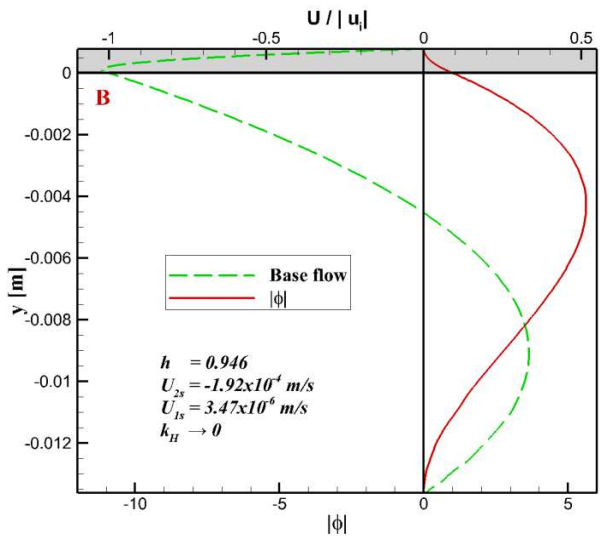
(c)

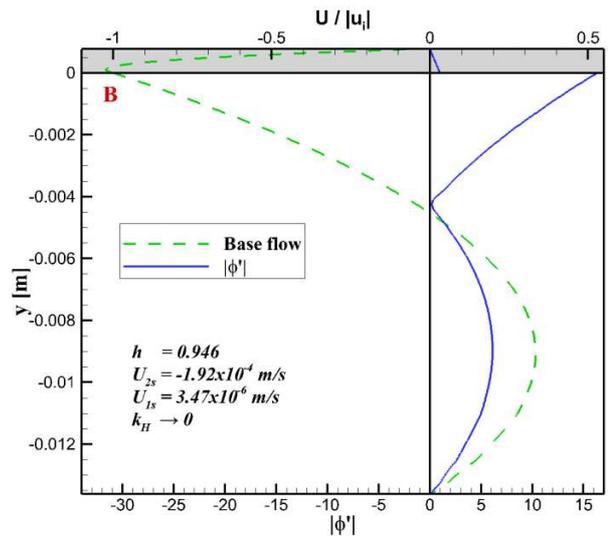
(d)

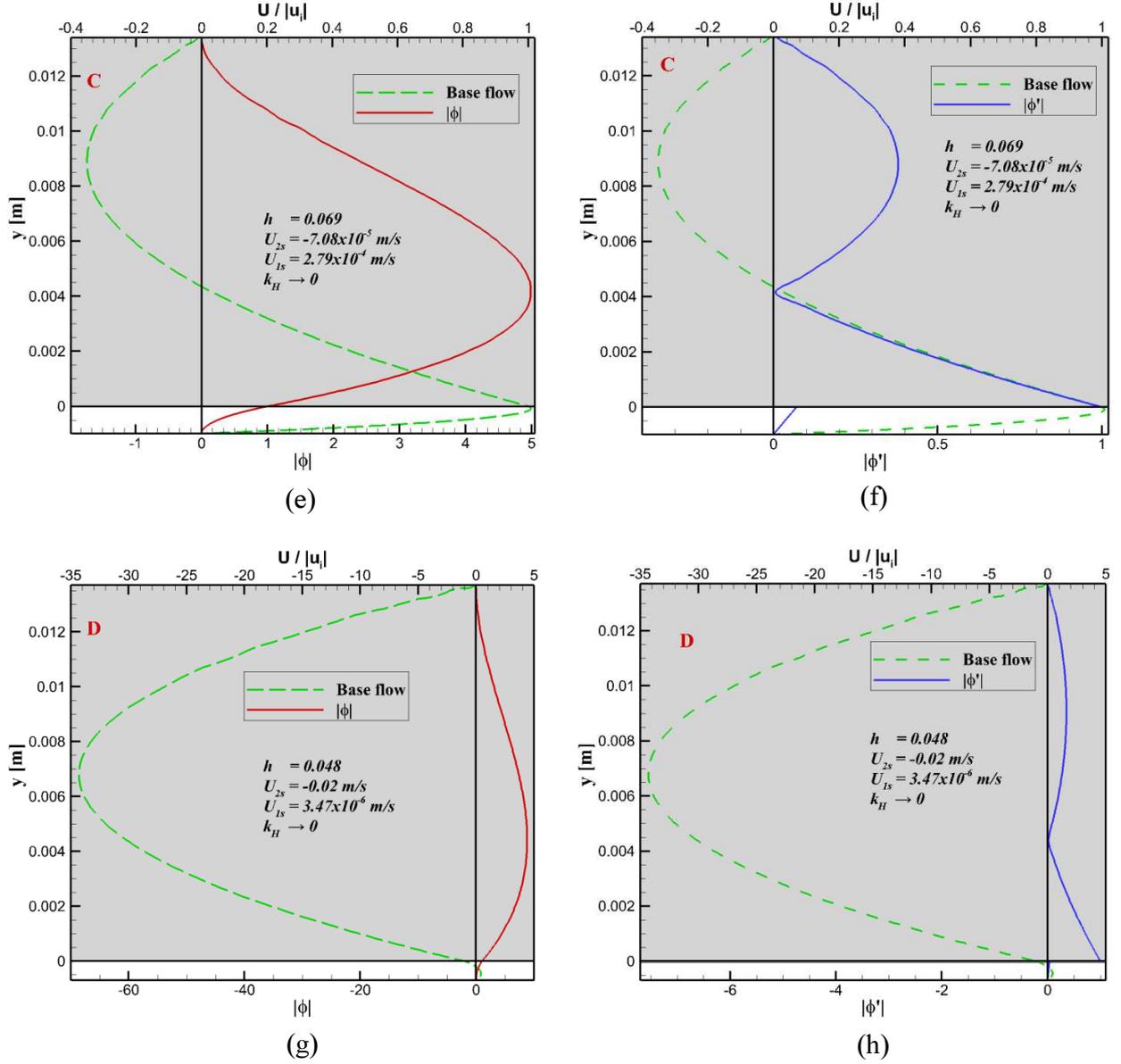

FIG. 7. Amplitudes of the critical perturbations of the stream function ((a), (c), (e), and (g): eigenfunction $|\phi|$, red solid lines) and its derivative ((b), (d), (f), and (h): $|\phi'|$, blue solid lines), and base flow velocity profile ($U/|u_i|$ green dashed lines) for the HPD solution at points A and B and for the LPD solution at points C and D (see Figure 4); y < 0 – heavy phase; y > 0 (shaded region) – light phase.

Further insight can be obtained by examining the two-dimensional contours of the real part of the stream function critical perturbations (Eq. (14)) at a particular time (e.g., $t = 0$), given by

$$\operatorname{Re}(\psi_j) = \operatorname{Re}(\phi_j(y)e^{ikx}) = \operatorname{Re}(\phi_j(y)) \cdot \cos(kx) - \operatorname{Im}(\phi_j(y)) \cdot \sin(kx). \qquad (24)$$

It is worth recalling that the amplitude cannot be determined in the framework of linear stability analysis. To observe the flow pattern of the critically perturbed flow, the critical perturbation with a small amplitude is superimposed on the base flow stream function (the latter is normalized to be unity at the interface). The appropriate value of the amplitude at the interface $|\phi(0)|$ should be found in each particular case to assure that the visualized flow field is consistent with the linear analysis. Therefore, the value $|\phi(0)|$ is chosen to be sufficiently small so that the disturbed (sinusoidal) interface coincides with a streamline in the frame of reference moving with a wave, in which the flow is steady. For relatively high amplitudes the streamlines are no longer sinusoidal, and some of them cross the interface, which is physically incorrect. The consequences of using appropriate and inappropriate values of the factor $|\phi(0)|$ are demonstrated in Figure 8(a) and (b), respectively.

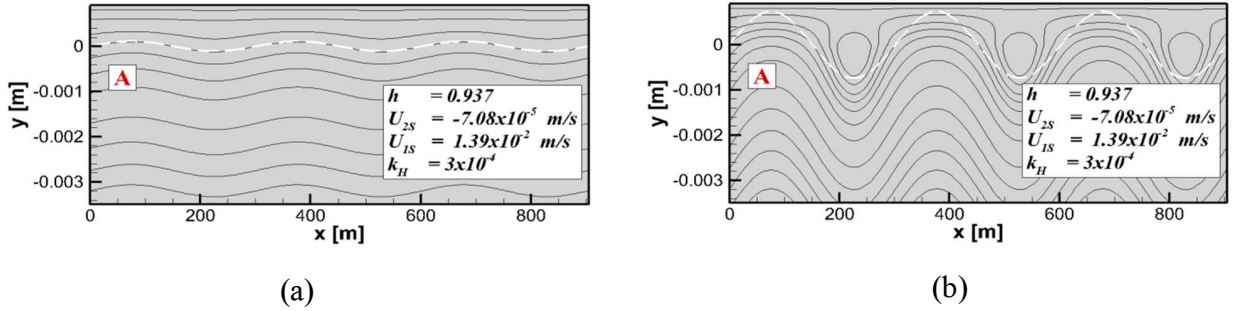

FIG. 8. Streamlines of the critically perturbed flow (base flow + scaled perturbation, black solid lines) and the corresponding disturbed interface (white dashed line) in the reference frame moving with the wave speed for the HPD solution at point A (see Figures 4, 7): (a) $|\phi(0)|$ = 1.5 (appropriate), (b) $|\phi(0)|$ = 10 (inappropriate); $0 \leq x \leq 3l_{wave}$ (only part of the lower layer adjacent to the interface is shown).

Figure 9 shows the 2D contours of the critical perturbations ($\text{Re}(\psi)$, left-hand side figures) and of the critically perturbed flow ($\text{Re}(\psi_{flow})$, right-hand side figures) in the stationary frame of reference for conditions corresponding to points **A-D** in Figures 4 and 7. As observed in Figures 9 (a), (c), (e), and (g), the perturbations are in phase across the entire flow domain of the countercurrent flows. In all tested cases the perturbations consist of pairs of antisymmetric vortices (of $l_{wave}/2$ width) with their core, i.e., the maximum of the stream function perturbations, located in the bulk of the thick layer. Although infinitely long waves $(k \to 0, l_{wave} \to \infty)$ are the critical perturbations in countercurrent flow, in practical applications the channel is of a finite length. Hence, in such cases a perturbation of a shorter wavelength is selected, whose growth is still close to zero (near critical condition), and the shape of streamlines and disturbed interface is similar to those of longer wave perturbations, including $k \to 0$ (e.g., $k_H = 0.01$ for point C, see Figure 5(a)). In this respect, Figure 7 corresponds also to the selected finite wavelength perturbations.

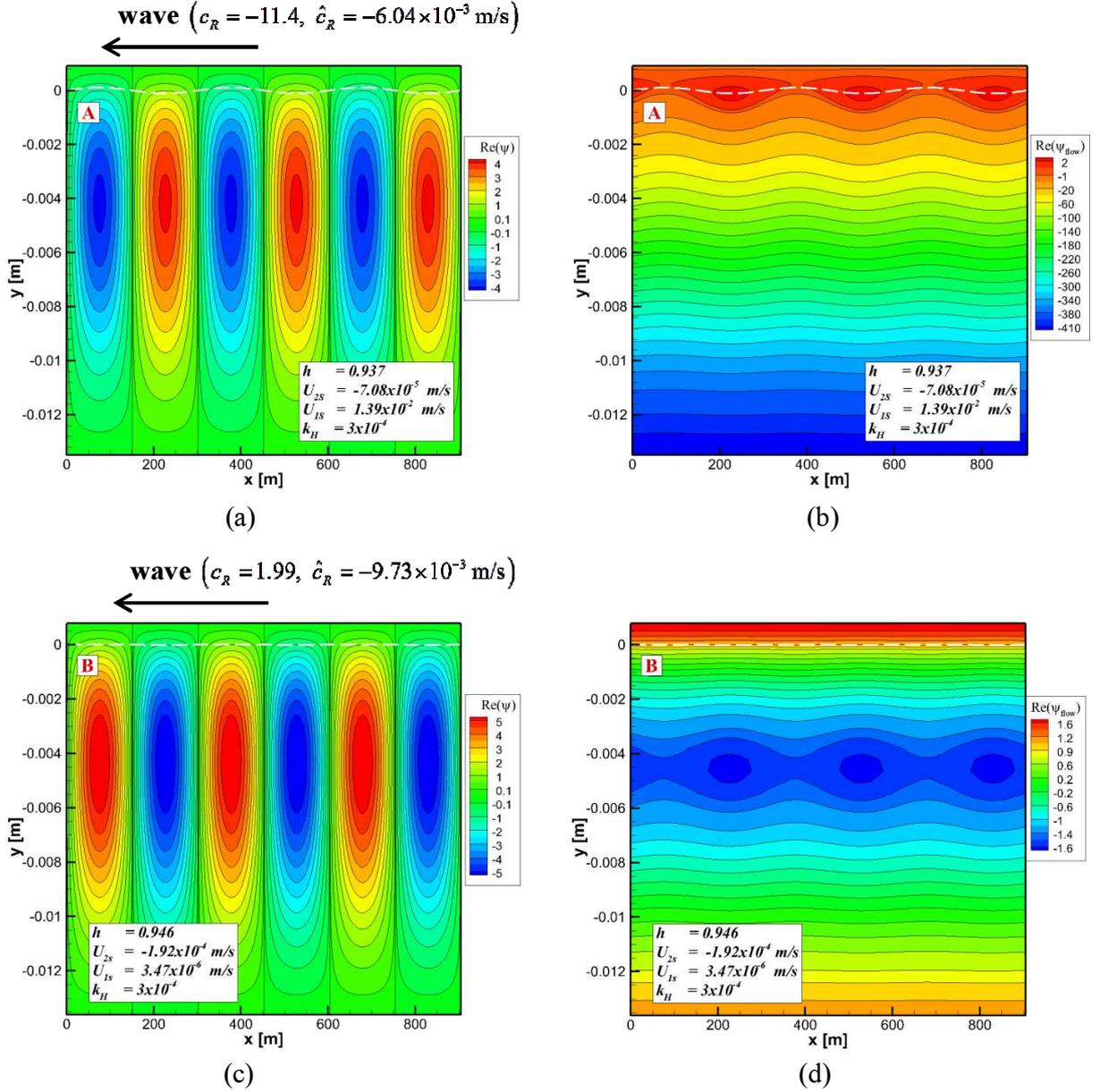

FIG. 9. Contours of the critical perturbations of the stream function ($\text{Re}(\psi)$, (a), (c), (e), and (g)) and of the critically perturbed flow (base flow + perturbation, $\text{Re}(\psi_{flow})$, (b), (d), (f), and (h), white dashed line is the corresponding disturbed interface) for the HPD solution at points A and B and for the LPD solution at points C and D (see Figure 4, 7); y < 0 – heavy phase; y > 0 – light phase; $0 \leq x \leq 3 l_{wave}$.

For the LPD configurations (**C**, **D**) the perturbations propagate downward in the direction of the thin heavy phase layer flow, while for the HPD cases (**A**, **B**) the perturbations propagate upward in the direction of the light phase thin layer flow. The direction of the wave propagation appears to be independent of the interface velocity direction ($u_i > 0$ in cases **A, C** and $u_i < 0$ in cases **B, D**). Although the stream function perturbations are similar, the perturbed flow differs for each case, owing to different base flow velocity profile. With no backflow in the dominating phase (points **A** and **D**), the waves with maximal amplitudes are observed near the interface. This results in deformation of the interface and generation of circulation cells, whose centers lie on the line of zero base state velocity (in the thin layer, see Figure 7). In these cases the thin layer is occupied by circulation cells, which may be considered as precursors for slugs. Note that the flow is not steady with respect to the stationary frame of reference (which is used in the present study), and the (moving) interface in a particular moment of time may not coincide with a streamline (e.g., see Figure 9(b)).

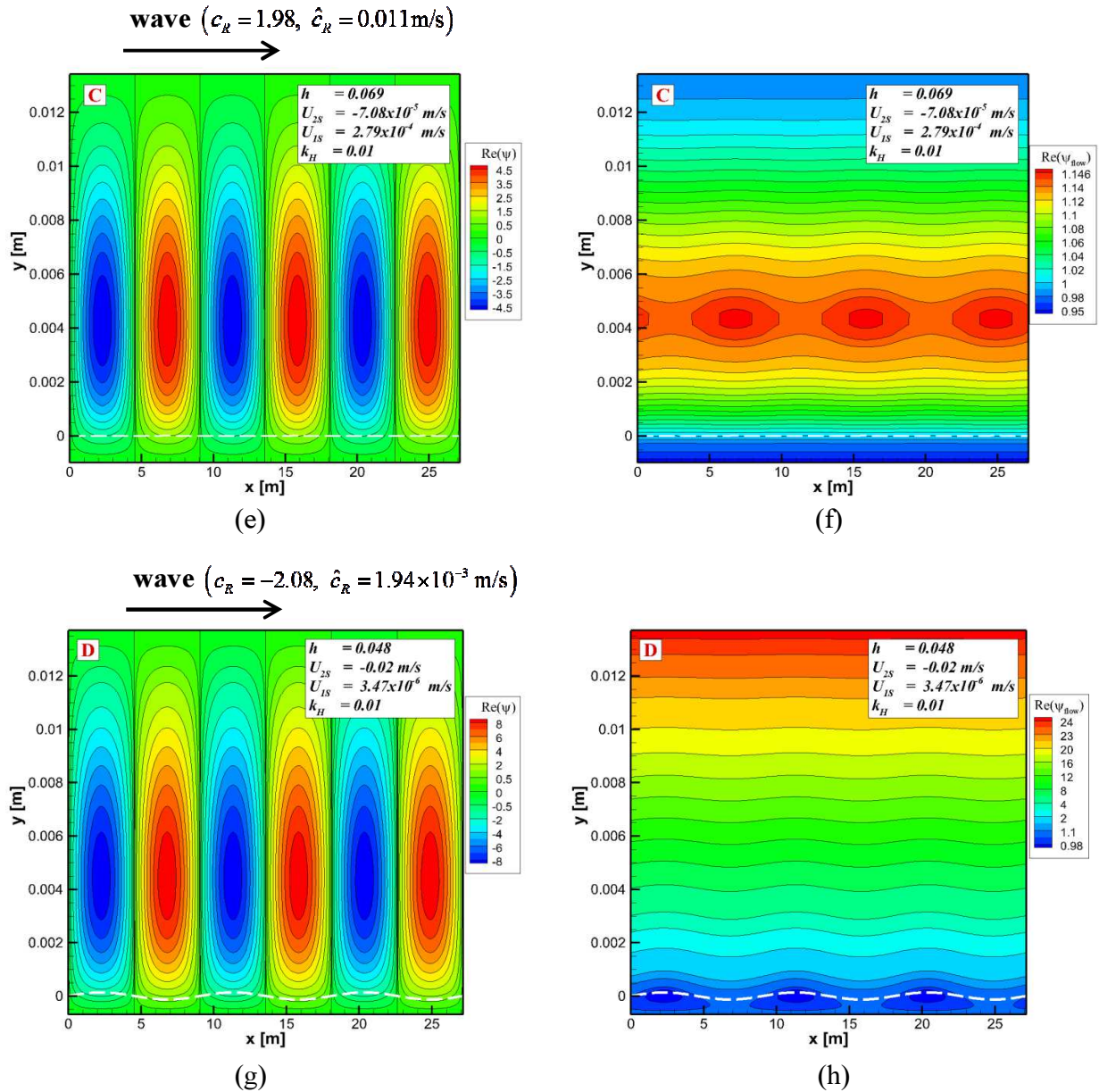

FIG. 9. (*Continued.*)

Another perturbed flow configuration is demonstrated for points **B** and **C** (Figures 9(d) and (f) respectively), where the large region of backflow exists in the base state. The largest amplitude waves are observed in the dominating phase leading to generation of circulation cells in its bulk (centered at the location of the zero base state velocity). The interface stays almost undeformed (flat) for these conditions. The similarity in the patterns of the critical perturbations observed in the cases of the LPD and HPD configurations is reasonable in view of the small differences in the fluids properties (i.e., the density and viscosity ratios are close to 1).

As shown in Figure 5, upon crossing the stability boundary, perturbations of a finite range of wavenumbers near the critical one become unstable. The obtained results show that the most amplified perturbation in near critical conditions (in the unstable region) is of a similar pattern to that of the critical perturbation, even though its wavenumber may be different. This holds, for example, in the case of the most amplified perturbation at point $C_2$ $\left(k_H = 0.22\right.$, Figure 5(a)), as well as for the spectrum of amplified modes at $A_2$ (up to $k_H$ about 0.005, Figure 5(b)). Therefore, inspection of the critically perturbed flow (i.e., on the neutral stability curve), may imply on the expected flow pattern that evolves in the unstable region. However, one should bear in mind that the linear analysis may be irrelevant in the unstable region (at conditions far from the stability boundary), where mode coalescence and other non-linear effects presumably take place.

Increasing the inclination angle results in shrinkage of the stable regions (see Figure 10 for $\beta = 26°$). These results are in accordance with the experimental findings in pipe flow (Zamir, 2003), where both possible countercurrent configurations with a smooth interface were observed provided the flow rate of one of the phases was sufficiently small. The predicted critical heavy (light) phase superficial velocity for the HPD (LPD) solutions for low flow rate of the light (heavy) phase is also in close agreement with the experimental values. At higher flow rates $\left(-U_{2S}, U_{1S} \geq 0.001 \text{m/s}\right)$, where both configurations are predicted to be unstable, wavy interface was observed. The maximal possible inclination of the channel where stratified smooth flow is still obtainable was found to be about 45° both by the stability analysis and in the experiments (see Ullmann et al, 2003a), beyond which the flow becomes unstable for all flow rates for the tested fluids. In case the heavy phase is less viscous than the light phase $(m = 0.5)$ the stable regions for the LPD and HPD configurations are reduced (see Figure 11), while the flooding curve is at higher flow rates.

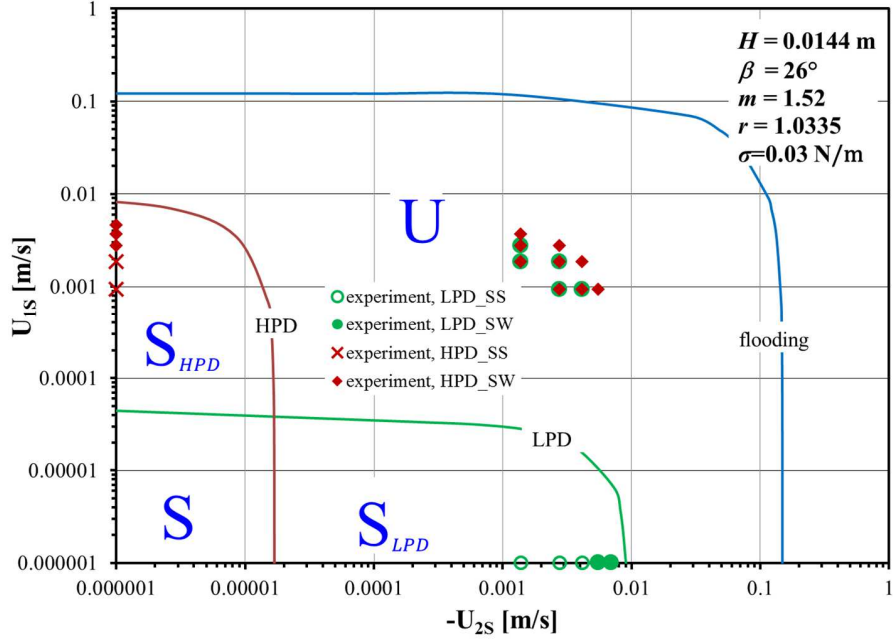

FIG. 10. Stability boundaries (solid lines) and experimental data (symbols: SS – stratified smooth, SW – stratified wavy pattern; points on the axes correspond to the lowest flow rates considered in the experiments; Zamir, 2003) for countercurrent liquid-liquid flow $(\beta = 26°)$.

Air-water countercurrent flow can be realized for a large range of air flow rates at shallow inclinations. However, the present analysis reveals that only the solution with very low holdup (LPD) is stable (see Figure 12). It was found that the stability of the LPD solution in the channel of 2 cm height is determined by long-wave perturbations. The stable region shrinks with increasing the inclination. Also in the channel of 5 cm height, long-wave analysis can be conveniently applied for predicting the neutral stability curve with sufficient precision, although there is a section of the stability boundary where the critical disturbances correspond to short-wave perturbations. Nonetheless, since the holdup is very low $(h \approx 0.01)$ for those conditions, the difference between the exact and long-wave boundaries is negligible. It is worth noting that the instability of the HPD configuration does not negate the possibility of obtaining two different countercurrent flow configurations (for the same operational conditions). It implies that only the LPD solution will result in stratified flow, while the growing interfacial waves in the HPD configuration will most probably result in bubbly or slug flow (see also Ullmann et al., 2003a).

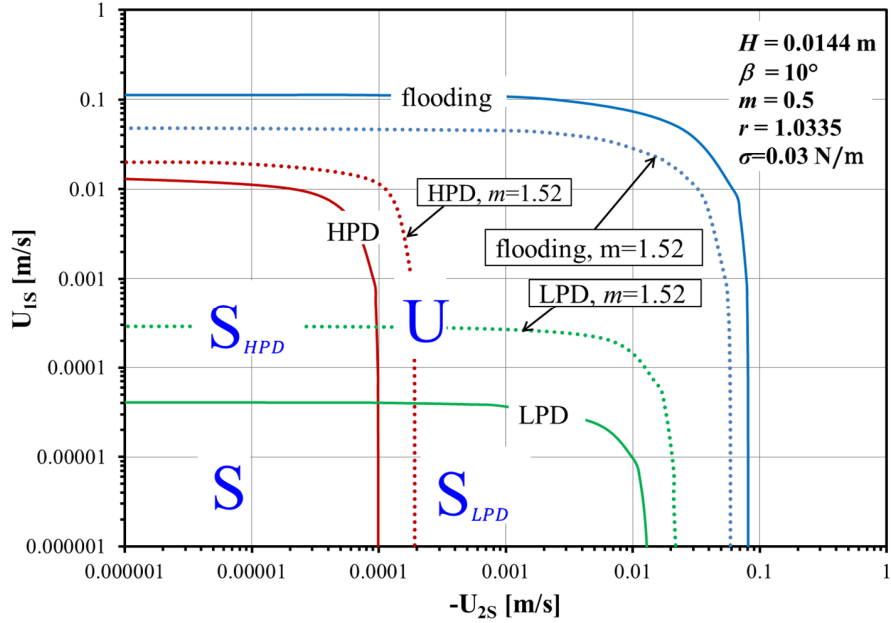

FIG. 11. Stability boundaries for countercurrent liquid-liquid flow for different viscosity ratios.

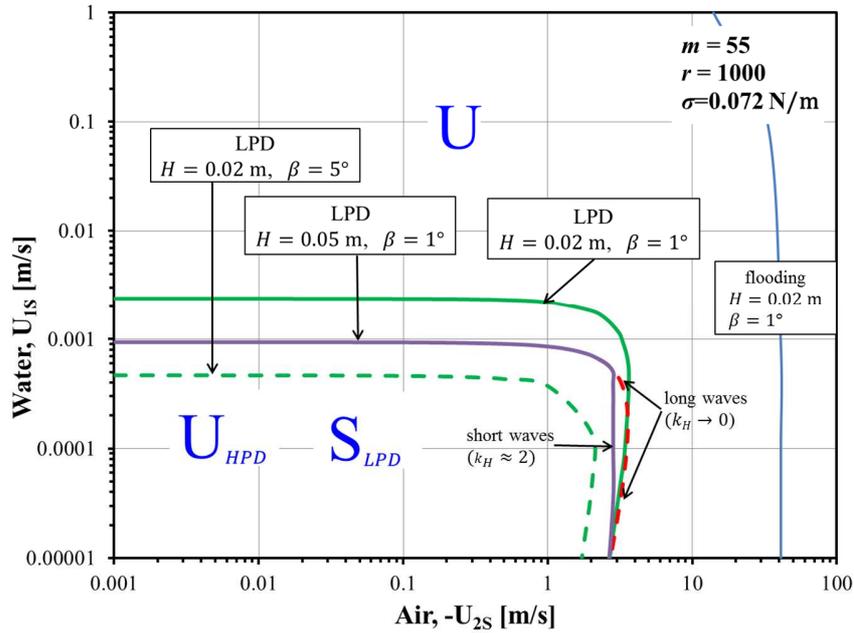

FIG. 12. Stability boundaries with respect to all wavenumber perturbations for countercurrent air-water flow of different channel heights ($H = 0.02, 0.05\,\text{m}$) and inclinations ($\beta = 1°, 5°$) and long-wave stability boundary for $H = 0.05$, $\beta = 1°$ (red dashed line). The flooding curves for $H = 0.05$, $\beta = 1°$ and $H = 0.02$, $\beta = 5°$ are at higher flow rates (out of the graph range). The upper solution for the holdup (HPD) is always unstable for the given parameters.

**B. Concurrent flows**

It is well established fact that multiple steady state configurations (i.e., three solutions for the holdup) in upward inclined gas-liquid flows are encountered for low liquid and high gas flow rates (Landman, 1991, Barnea and Taitel, 1992, Ullmann et al., 2003). This is a region of practical importance, where stratified smooth and wavy flows are observed in experiments (e.g., Barnea et al., 1980, and discussion in Ullmann et al., 2003b). The location of the triple solution region on the air-water flow pattern map depends on the channel inclination and height. In this region, in addition to the single low holdup solution (obtained at high gas flow rates), two solutions of the medium and high holdups are obtained. The larger is the channel height (and/or angle of inclination) the larger are the air flow rates corresponding to multiple solutions for the holdup. For particular angle $\beta = 0.1°$ the boundaries of triple solution regions for different channel heights are shown in Figure 13. From the practical point of view of relevant air flow rates, the triple solution regions of interest correspond to channels heights ranging from 1 cm and up to 10 cm.

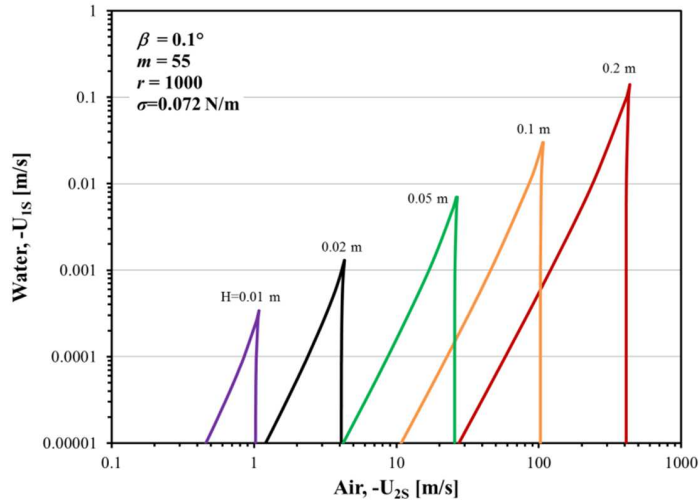

FIG. 13. Triple solution regions for slightly upward inclined $(\beta = 0.1°)$ air-water flow for different channel heights.

A detailed stability analysis is conducted for the case of concurrent air-water flow in a slightly upward inclined 2 cm channel. The long-wave stability of this system was studied by Kushnir et al. (2014). The stability map that considers all wavenumber perturbations is presented in Figure 14. In concurrent up-flow, the gravity force slows down the heavy phase (water), and consequently the holdup is much higher than in horizontal flow. Even at low water flow rates and shallow inclination the water occupies most of the flow cross section. Consequently, at low gas flow rates, where only a single solution of high holdup exists, the critical air velocity is reduced drastically in comparison to the critical value in horizontal channels (Barmak et al., 2016) and the long-wave perturbation is found to be the critical mode. However, the critical superficial water velocity for low air flow rates is close to that obtained in a horizontal channel. At low air flow rates, the instability is mainly associated with destabilization of the thick water layer and is determined by short and intermediate wavelength disturbances (the critical disturbance wavenumbers, normalized by the channel height, $k_H = 2\pi H / l_{wave}$, are depicted in Figure 14 on the stability boundary). This stable region is however limited to very low air flow rates (and gas void fractions), which are out of the range of experimental studies, and most probably will be occupied by bubbly flow due to capillary instability at the feeding device.

With increasing the gas flow rate, as long as a single high holdup is obtained, the flow remains unstable. Stable stratified flow is predicted (and experimentally observed, e.g., see Figures 2 in Barnea et al., 1980) at higher gas flow rates upon entering the triple solution region, where additional two solutions of lower holdups are obtained. For example, in the experiments of Barnea et al. (1980) in a 2.5 cm pipe of a 0.25° upward inclination, stratified flow (with a smooth or a wavy interface) was observed at $U_{2S} \approx 2 \div 11 \,[\text{m/s}]$ and $U_{1S}$ is up to $0.03 \,[\text{m/s}]$, i.e., in the range of flow rates corresponding to the triple solution region and its vicinity.

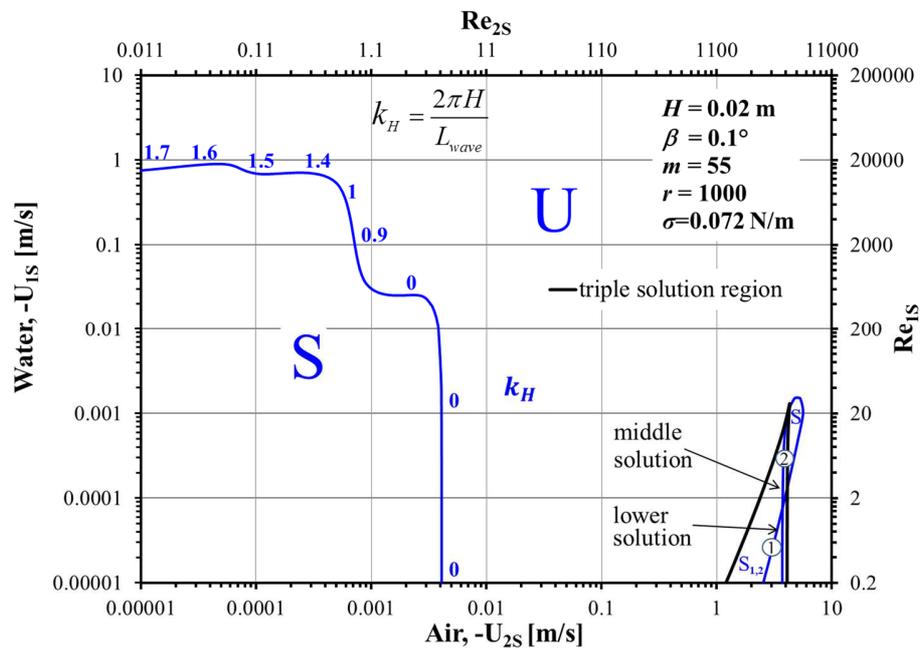

FIG. 14. Stability boundaries for slightly upward inclined air-water flow.

Kushnir et al. (2014) revealed that in slightly inclined channel all three holdup solutions in the multiple solutions region can be stable to long-wave disturbances. Therefore, different (smooth) stratified flow configurations may be obtained depending on the history of the flow (i.e., hysteresis phenomenon). According to those results (reproduced in Figure 15(a) where the triple solution region is enlarged), the middle solution is stable within the entire triple solutions region. The lower solution (line 1) is stable in regions **a** and **b**, and the upper solution (line 3) is stable in regions **b** and **c**. Accordingly, in region **d** only the middle solution is stable. Consideration of all wavelength perturbations changes the picture (see Figure 15(b)). The upper solution becomes unstable in the entire triple solution region. The lower solution (line 1) is stable in the **a** and **b** regions, and its stable domain extends outside the triple solution region to higher gas rates, where the water holdup is still low. The middle solution (line 2) is stable in regions **a** and **d**. In region **c** none of the solutions are stable.

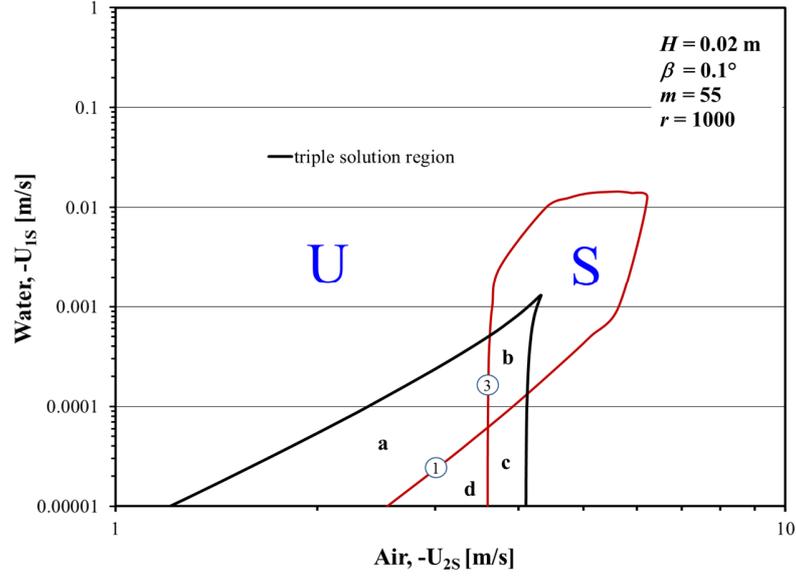

(a)

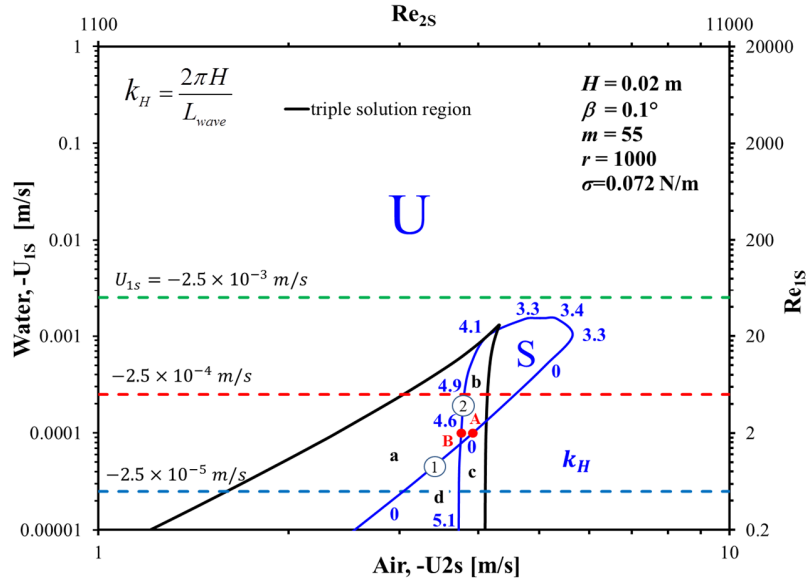

(b)

FIG. 15. (a). Long-wave stability boundaries in the triple solution region of slightly upward inclined air-water flow (from Kushnir et al., 2014):  **a** – lower and middle solutions are stable;  **b** – all three solutions are stable;  **c** – middle and upper solutions are stable;  **d** – only the middle solution is stable. Lower, middle and upper solutions - lines 1, 2, and 3, respectively. (b). All-wave perturbations stability boundaries in the triple solution region of slightly upward inclined air-water flow:  **a** – lower and middle solutions are stable;  **b** – only the lower solution is stable;  **c** – all three solutions are unstable;  **d** – only the middle solution is stable.

The above results can be made clearer when represented in the form of holdup curves at various fixed water flow rates. Such holdup curves are plotted in Figure 16, where dashed lines represent unstable conditions. For $U_{1s} = -2.5 \cdot 10^{-5}\,\text{m/s}$ the single high holdup solution obtained at relatively low air flow rates is unstable. Upon increasing the air flow rate and crossing the multiple region boundary, two additional solution of lower holdups (low and middle solutions) are obtained which are stable. With further increase of the air flow rate, the lower solution becomes unstable, while the middle solution becomes unstable only at higher air flow rates. At larger (in absolute value) water superficial velocity, $U_{1s} = -2.5 \cdot 10^{-4}\,\text{m/s}$, the middle holdup solution becomes unstable at lower air flow rates than the low holdup solution, which remains stable throughout the entire multi-holdups region (and stays stable for a range of air flow rates outside the triple solution region). In contrast to long-wave analysis it was found that for water rates higher than the maximum allowing for multiple solutions ($-U_{1s} > 1.5 \cdot 10^{-3}\,\text{m/s}$, e.g., $U_{1s} = -2.5 \cdot 10^{-3}\,\text{m/s}$) the smooth stratified flow configuration is unstable for the whole range of air flow rates (exceeding $0.004\,\text{m/s}$, see Figures 14, 15(b)). As shown in Figure 16, in the vicinity of the triple solution boundary, a small decrease of the air flow rate is accompanied by a sharp increase of the holdup (and the hydrostatic pressure drop). It can be concluded that even for slight upward inclination only a thin film of water can be stable for high air flow rates.

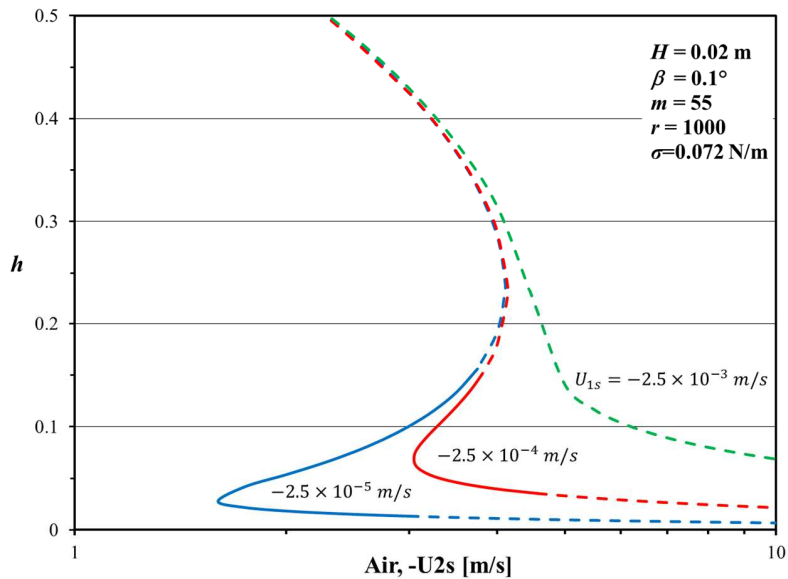

FIG. 16. Holdup curves for constant water flow rate in slightly upward inclined flow. Dashed lines represent unstable conditions.

The amplitudes of the stream function corresponding to the critical perturbations obtained for the lowest holdup solution at point **A** (very thin layer of the water), and those obtained for the middle solution with the same water superficial velocity at point **B** (marked in Figure 15(b)), are shown in Figures 17(a) and (c) respectively. In both cases the heavy phase (water) is dragged upward by the fast flowing air, and the only solution without backflow is that of the lowest holdup. At point **B**, the velocity profile shows a small region of backflow of the water near the lower wall. Due to the high viscosity ratio $(m=55)$, the interface behaves similarly to a rigid wall with respect to the air flow. At point **A** the instability is triggered in the bulk of the air layer (Figure 17(a)), implying a shear mode instability, however with a long wavelength, whereas the largest value of the streamwise velocity disturbance (i.e., eigenfunction derivative) is located at the interface (see Figure 17(b)). This disturbance is similar to the one reported for the horizontal air-water flow with low holdup (Barmak et al., 2016). On the other hand, the middle solution becomes unstable owing to a short wave with a maximal disturbance in the air close to the interface, and an additional peak in the water at the interface (see Figures 17(c) and (d)). This implies an interfacial mode of instability. The same characteristics of the critical perturbations can be attributed to the conditions all along the neutral stability curves inside the triple solution region. These results further substantiate the findings obtained in horizontal flows, which indicate that there is no definite correlation between the type of instability and the perturbation wavelength (i.e., shear mode can be associated also with long waves, and interfacial mode with short-wave perturbations).

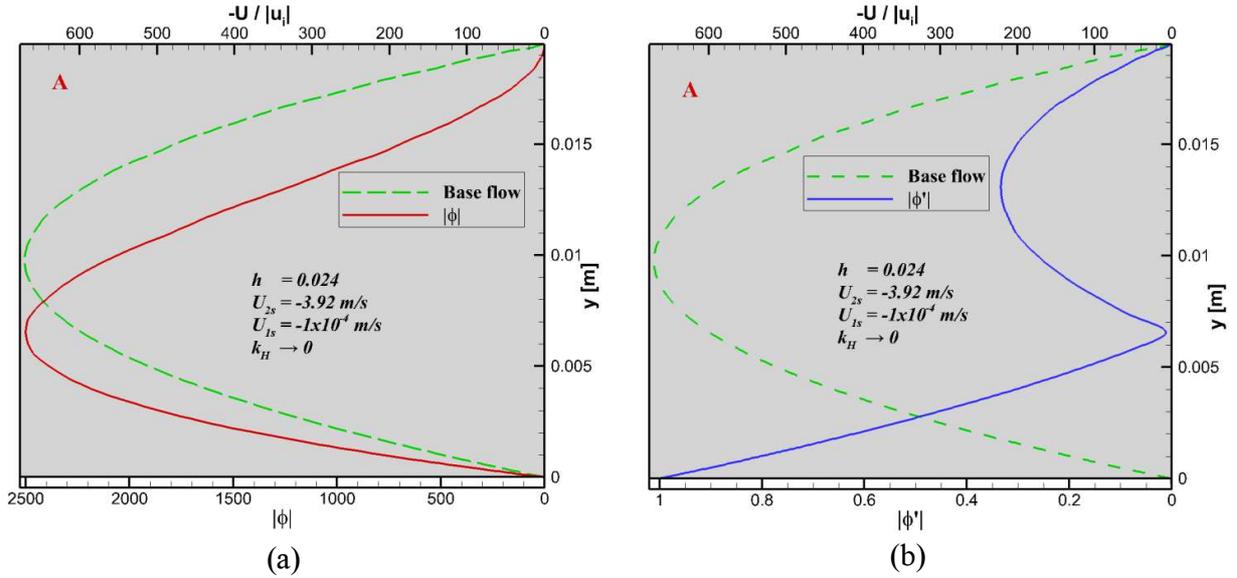

(a)     (b)

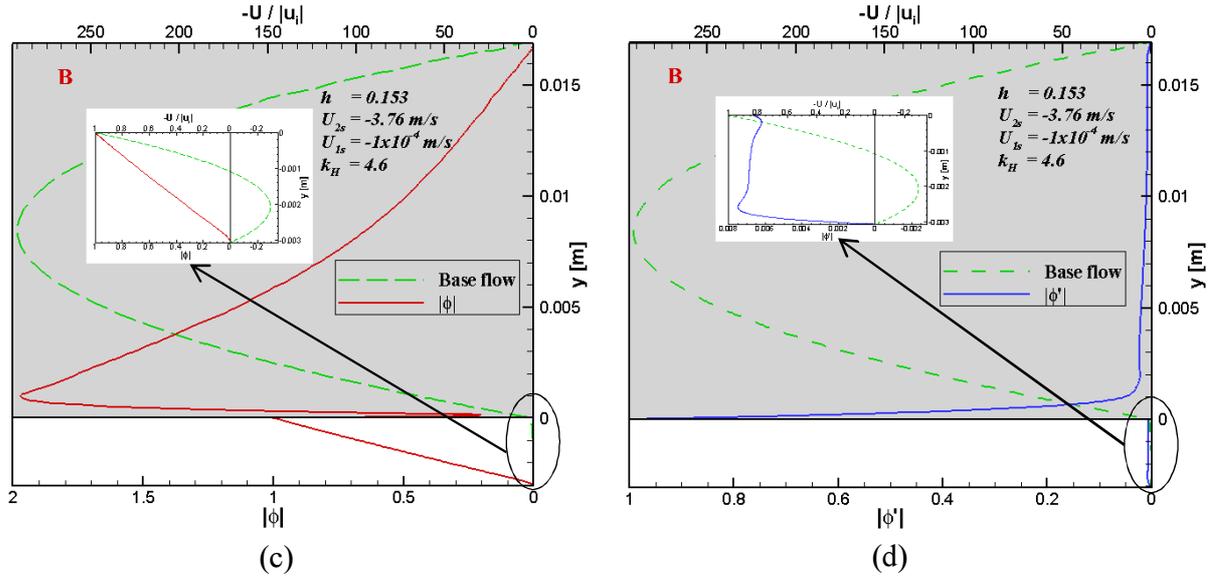

FIG. 17. Amplitudes of the critical perturbations of the stream function ((a), (c), (e), and (g); eigenfunction $|\phi|$, red solid lines) and its derivative ((b), (d), (f), and (h); $|\phi'|$, blue solid lines), and base flow velocity profile ($U/|u_i|$, green dashed lines) at points A (lower solution) and B (middle solution; see Figure 13 (b)); $y < 0$ – water; $y > 0$ (shaded region) – air.

    Another important issue that can be brought up for consideration is the possibility of the so-called viscous Kelvin-Helmholtz (K-H) instability in the studied cases. This type of instability is commonly considered as the main destabilizing mechanism in the framework of the Two-Fluid model approach to analyze the stability of the flow. Basically, it is a viscous modification of the classical (inviscid) K-H instability that attributes the growth of interfacial perturbations to the destabilizing effect of inertia over the stabilizing effect of buoyancy. Obviously, these two effects are included in the present analysis (their ratio is represented by Froude number). In terms of the perturbation pattern, the K-H instability mode is typically limited to the vicinity of the interface (maximum value at the interface) and therefore is associated with an interfacial mode of instability. Note that the original (inviscid) Kelvin-Helmholtz instability was not included in the classification scheme of Boomkamp and Miesen (1996), who examined the different mechanisms by which energy transfer from the base flow to perturbations can take place. They argued that the possibility of the essentially inviscid Kelvin-Helmholtz instability is ruled out by the presence of viscous effects, however small.

    As mentioned before, in all considered cases of countercurrent flow and for the lower holdup solution in upward inclined flow, the critical perturbation was classified as a shear mode of instability. Therefore, the K-H instability is obviously not dominant in these cases. In this respect, a case of interest is the one where the critical perturbation is associated with an interfacial mode of instability. This can be observed for the critical perturbation at point B of the middle holdup solution in upward inclined flow (see Figure 17(c) and (d)). The relevance of the K-H mechanism can be quantified by calculating the terms representing the (destabilizing) inertia of each of the phases relative to the (stabilizing) gravity (for details see, e. g., Brauner and Moalem Maron, 1993, Kushnir et al., 2007), which should be of the order of 1 to indicate dominancy of the K-H instability:

$$J_1 = \left(\frac{\rho_1}{\rho_1 - \rho_2}\right)\frac{U_{1S}^2}{Hg\cos\beta}\frac{1}{h^3}\left[\left(\frac{\hat{c}_R}{\bar{U}_1} - 1\right)^2 + (\gamma_1 - 1)\left(1 - 2\frac{\hat{c}_R}{\bar{U}_1}\right) + \Delta\gamma_1\right], \quad (25)$$

$$J_2 = \left(\frac{\rho_2}{\rho_1 - \rho_2}\right)\frac{U_{2S}^2}{Hg\cos\beta}\frac{1}{(1-h)^3}\left[\left(\frac{\hat{c}_R}{\bar{U}_2} - 1\right)^2 + (\gamma_2 - 1)\left(1 - 2\frac{\hat{c}_R}{\bar{U}_2}\right) + \Delta\gamma_2\right]. \quad (26)$$

The $J$ terms are calculated for the steady (viscous) flow conditions represented by average phases velocities $\bar{U}_{1,2}$ and holdup $h$. The shape factors $\gamma_{1,2}$ (and their derivative terms, $\Delta\gamma_{1,2}$) represent the correction of the phases' inertia terms (when expressed in terms of the average phase velocity) and are determined based on the exact velocity profiles. In the framework of Two-Fluid model, where the only destabilizing (stabilizing) effect is due to the phases inertia (gravity), the sum of these $J$ terms should amount to 1 on the neutral stability boundary. It is further to be noted that in the Two-Fluid model the wave velocity $\hat{c}_R$ corresponds to a (kinematic) long wave. However, for the sake of evaluating the significance of the destabilizing effects of inertia, the wave velocity of the critical (no matter long or short) perturbation is utilized in Eqs. (25) and (26). Accordingly, the values of $J_1$ (inertia of the water layer) and $J_2$ (inertia of the air layer) obtained for the critical perturbation at point B are 1.12 and 0.1, respectively. This result suggests a water-dominated K-H mechanism of instability for the considered configuration of air-water stratified flow in upward inclined channel. Obviously, since in the current exact analysis all the stabilizing and destabilizing (linear) mechanisms are included, the sum of the above $J$ terms does not necessarily amount to 1 on the stability boundary. On the other hand, in the framework of the long-wave analysis, when $J_1$ and $J_2$ are calculated by using the velocity of a kinematic wave, one can find that the inertia of both phases (air and water) is insignificant in the entire triple solution region. Following this approach it can be reassured that the K-H instability does not arise in the other considered cases (with shear mode of instability), since the terms $J_{1,2}$ are small compared to 1, implying that other mechanisms are responsible for instability of the flow.

The 2D contours of the stream function corresponding to the above critical perturbations (from Figure 17), normalized by the absolute value of their amplitude at the interface, are presented in Figures 18(a) and (c). At point **A** the perturbation vortices occupy the gas layer, and are in phase across the flow. A superposition of this critical perturbation with the base state reveals that the largest waves arise in the vicinity of the interface, which leads to deformation of the interface (see Figure 18(b)). The critical perturbation for the middle holdup solution (point **B**) is more complicated as two perturbation vortices with a phase shift in the flow direction are superimposed near the interface with the core (maximal perturbation) of one of them in the light phase and of the other – at the interface (see Figure 18(c)). The discontinuity in the derivative of the stream function perturbation at the interface (observed in Figure 17(d)) leads to abrupt bends in the streamlines. The streamlines of the neutrally stable perturbed flow attain maximal wave amplitudes in the vicinity of the interface, while the circulation cells formed in the water layer owing to the water backflow (downward flow) adjacent to the lower wall (see Figure 18(d)). In contrast to the other considered cases, at point **B** the short-wavelength (whereas long waves in the others) perturbation is the critical one, where in experiments even small amplitude interfacial waves are possibly detectable. The phase speed of the critical perturbation is negative in both of the cases studied (point **A** and **B**) therefore waves and circulation cells generated in the flow propagate upward in the downstream direction.

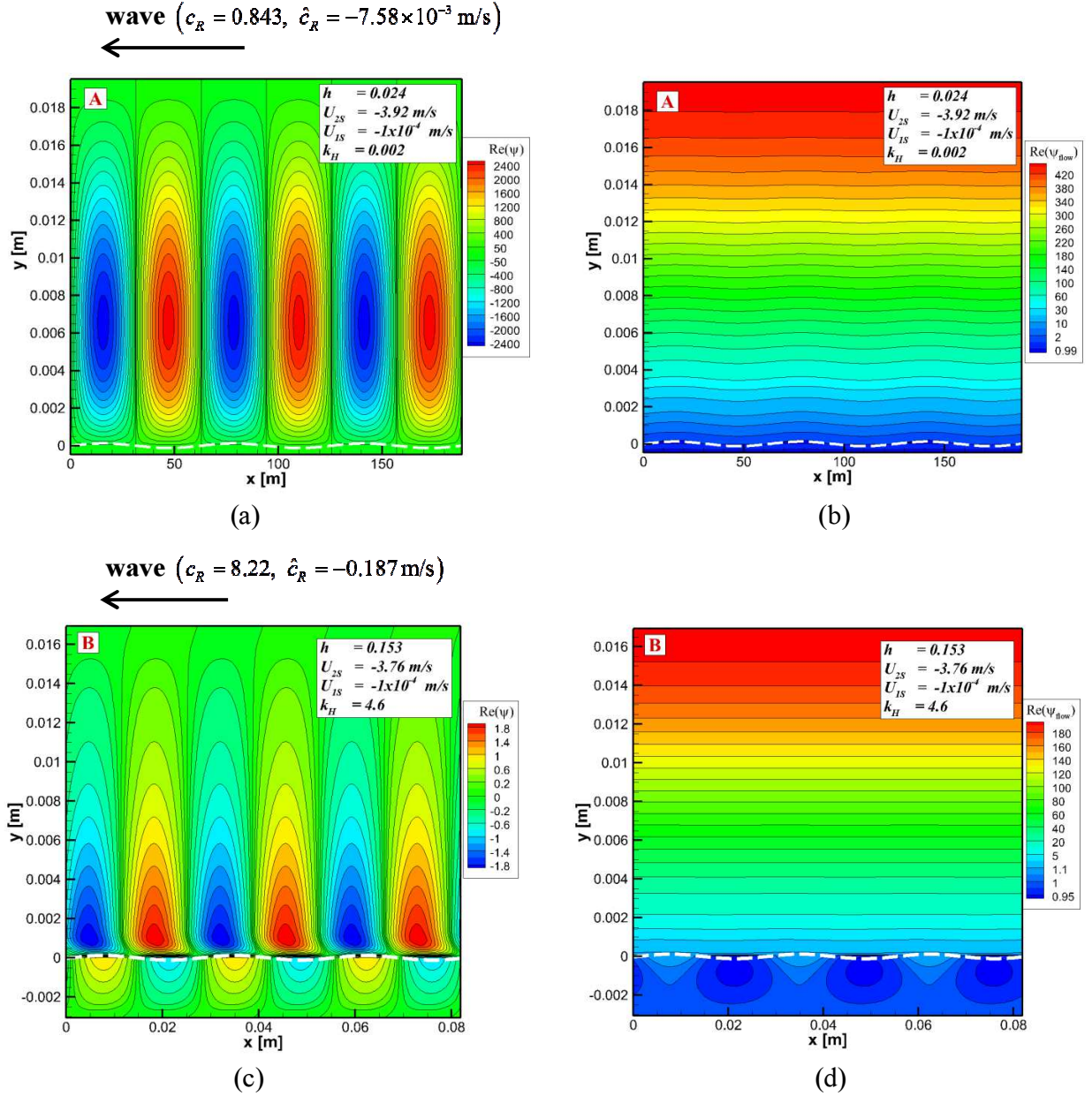

FIG. 18. Contours of the stream function of the critical perturbations ($\text{Re}(\psi)$, (a) and (c)) and of the critically perturbed flow (base flow + perturbation, $\text{Re}(\psi_{flow})$ (b) and (d), white dashed line is the corresponding disturbed interface) for the lower solution for the holdup at point A and for the middle solution at B (see Fig.13 (b)); $0 \le x \le 3 l_{wave}$. The $x$ coordinate is in the downward direction.

Stability analysis of slightly upward inclined flow reveals also that in the triple solution region the stratified flow is always unstable in the channels of 5 cm height and above. This result seems reasonable in light of the certain similarity between slightly upward inclined and horizontal flows for low water holdups, where the flow is dominated by the shear in the gas phase. On the one hand, with increasing the channel height, the triple solution region is shifted to higher gas flow rates (see Figure 13 ), and, on the other hand, the maximal air flow rate for maintaining a smooth interface decreases with the channel height due to short-wave instability (see Barmak et al., 2016). Depending on the holdup, either stratified wavy or slug flow can be expected in the region of multiple holdups in channels exceeding $\approx 5 \text{cm}$ in height.

The stability regions for air-water flows in $\beta = 1°$ and $5°$ downward inclined channels are shown in Figure 19. For concurrent downward flow, the triple solution region is obtained at high liquid flow rates and relatively low gas flow rates. As shown in Figure 19, downward inclined flow is stable only in the single solution region. The stable region is limited to significantly lower liquid flow rates (and holdups) compared to horizontal channel (see Barmak et al., 2016). In contrast to horizontal flow, the critical water superficial velocity results from a long-wave instability, which is the critical disturbance in this case all along the stability boundary. The resulting critical water superficial velocity is practically constant over a wide range of (low to medium) gas superficial velocities. The inclination of the channel results also in a lower critical gas superficial velocity (at low superficial water velocity), although it is much less sensitive to the inclination than the critical water superficial velocity. The obtained results can be validated by comparison with existing experimental data for slightly inclined downward flow (e.g., see Figures 3 and 5 in Barnea et al., 1982). The predicted region of smooth stratified flow is similar to that observed in air-water pipe flow, although the experimentally observed critical water superficial velocity is somewhat higher. This can be attributed to the finite length of the experimental facility, where very long waves may not exist. For higher liquid flow rates the triple solution region appears, but none of the solutions are stable, because of short-wave instabilities in the thick water layer. This is in contrast to the results obtained via long-wave stability analysis, where stable solutions are predicted in the triple solution region (Kushnir et al., 2014). Increase of the inclination angle to $5°$ results in a further shrinkage of the stable region, and the multiple solution region is shifted to higher flow rates, where all possible configurations are unstable for short-wave perturbations.

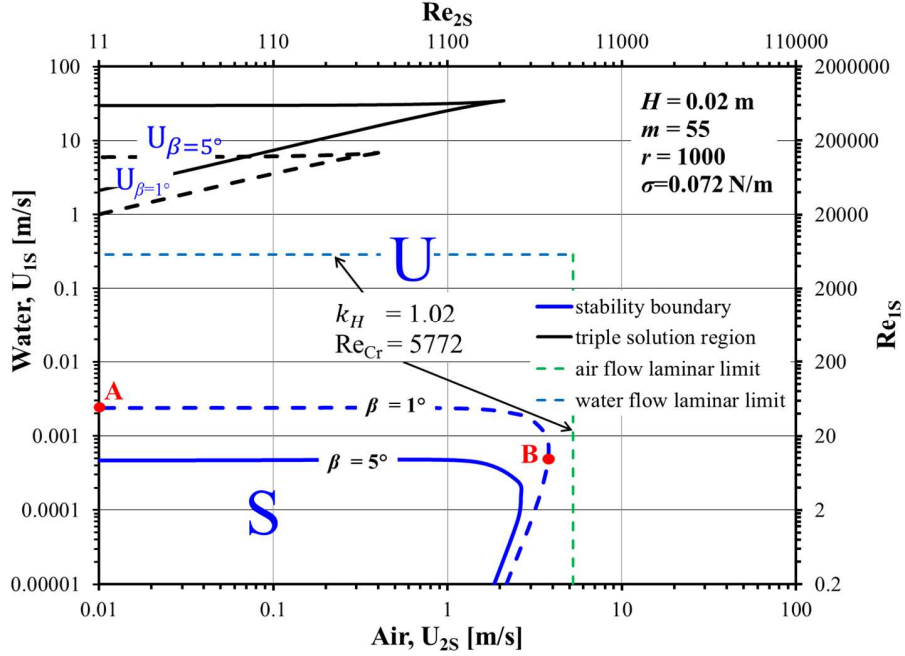

FIG. 19. Stability map for downward inclined air-water flow and air and water single-phase laminar limits.

The limits for laminar air and water single-phase flows are also depicted in Figure 19. The location of these lines does not depend on the inclination and is the same as for the corresponding horizontal flow, and determined by the critical Reynolds number ($\text{Re}_{Cr} = 5772$, critical wavenumber $k_H = 1.02$ (e.g., Orszag, 1971)). Although for horizontal air-water flow a stabilization effect (compared to single-phase water flow) was found at low air flow rates (see Barmak et al., 2016), downward inclined flow becomes unstable for lower flow rates than those that correspond to initiation of turbulence in single-phase flow.

Instability in downward flows is triggered by long-wave modes whose patterns correspond to the shear instability with the eigenfunction maximum located in the bulk of thick air layer (see Figure 20). The perturbation patterns are similar for two points **A** and **B** for $\beta = 1°$ downward inclined flow (marked in Figure 19). This similarity is despite the fact that the base state velocity profiles at point **A** has the largest velocity at the interface and (upward) backflow of the air occupies the zone adjacent to the upper wall, while the velocity distribution for point **B** is similar to that of the single phase air flow. It is significant to mention that similarly to the stability boundary of the lower solution in concurrent up-flow, also for downward flow, the dominant perturbations and the stability boundary for low water holdups and high air flow rates look similar to those of horizontal air-water flow. This is expected since under these conditions the flow is shear-dominated and gravity effects are mild. On the other hand, the entire stability boundary of the downward inclined flows resembles that obtained for the LPD configuration in countercurrent air-water flow (see Figure 12). Contours of the stream function of the perturbations and of the perturbed flow are demonstrated in Figure 21. The critical perturbations are in phase across the flow, and their phase speed is positive. For small air flow rates circulation cells (centered in the locus of zero base flow velocity) are generated in the critically (neutrally stable) perturbed flow in the bulk of the air phase, where the perturbation is maximal. For high air flow rates the streamlines of the perturbed flow are wavy in the upper layer with high amplitudes close to the interface, which results in its deformation.

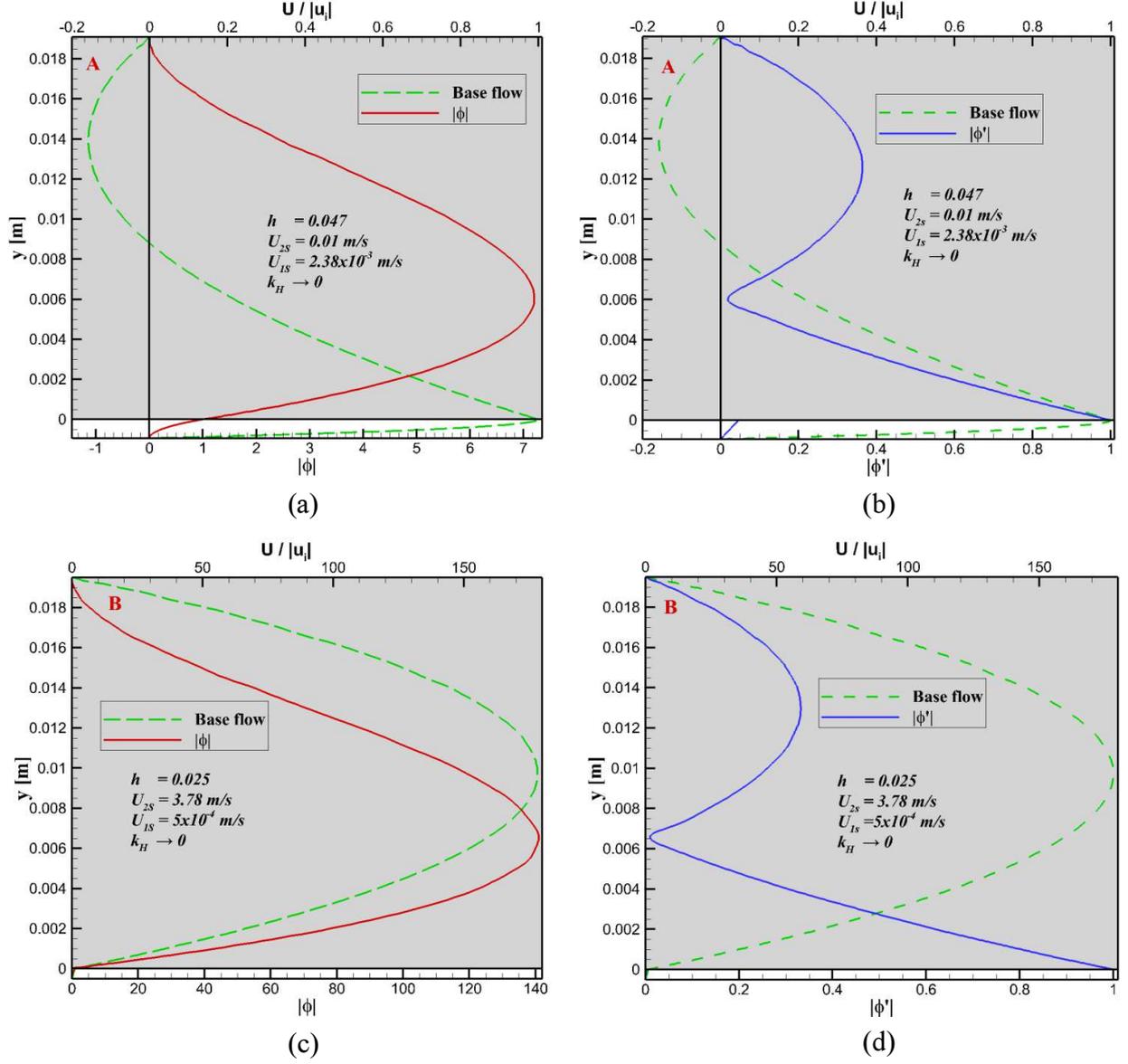

FIG. 20. Amplitudes of the critical perturbations of the stream function ((a) and (c); eigenfunction $|\phi|$, red solid lines) and its derivative ((b) and (d); $|\phi'|$, blue solid lines), and base flow velocity profile ($U/|u_i|$, green dashed lines) at points A and B (see Figure 19); $y < 0$ – water; $y > 0$ (shaded region) – air.

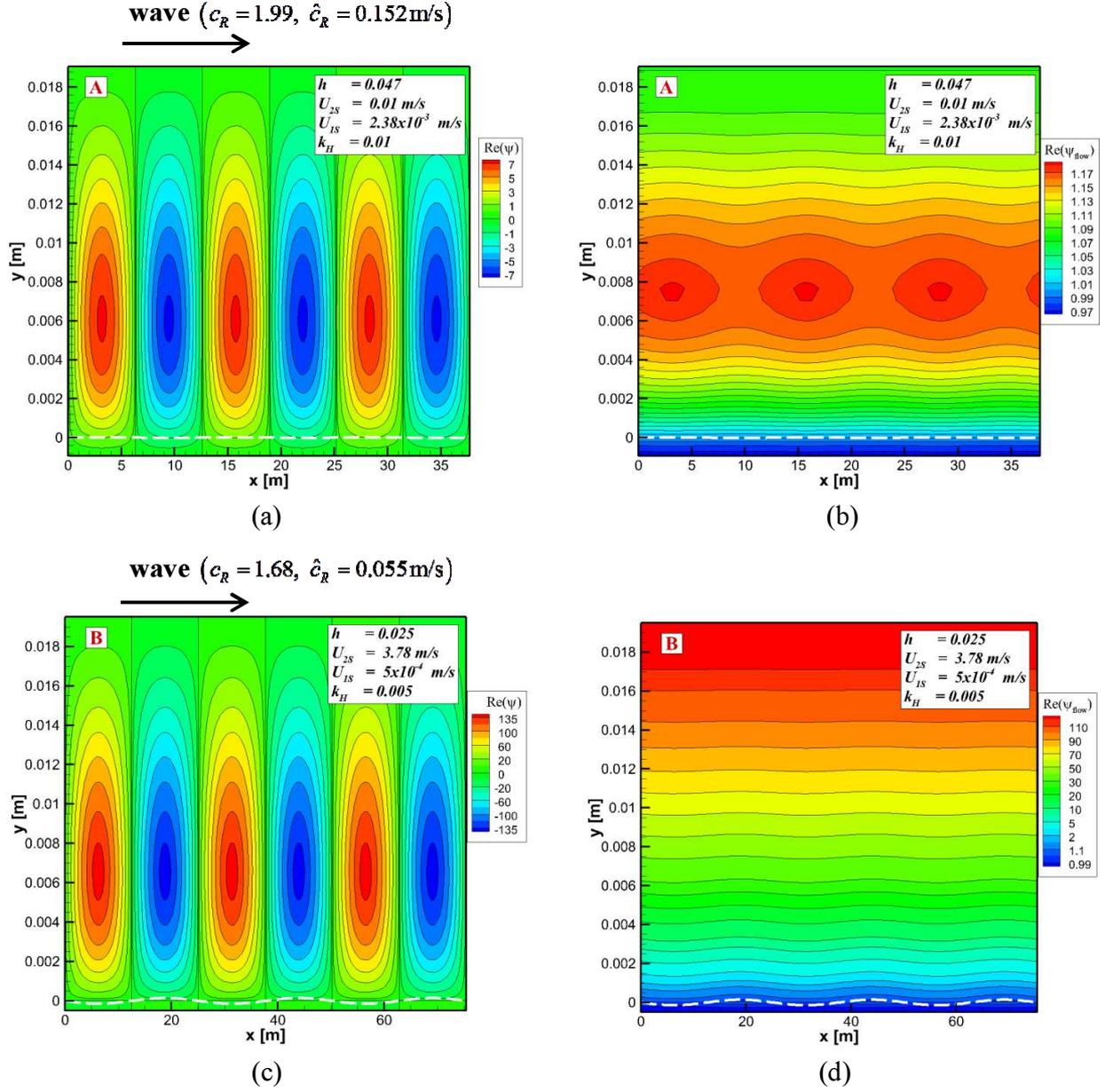

FIG. 21. Contours of the stream function of the critical perturbations ($\text{Re}(\psi)$, (a) and (c)) and of the critically perturbed flow (base flow + perturbation, $\text{Re}(\psi_{flow})$, (b) and (d)), white dashed line is the corresponding disturbed interface) for downward inclined air-water flow at points A and B (see Figures 19, 20); $0 \leq x \leq 3l_{wave}$.

## VI. CONCLUSIONS

The linear stability of laminar countercurrent and concurrent gas-liquid and liquid-liquid plane-parallel flows between two plane boundaries was studied. The analysis is based on the well-known Orr-Sommerfeld eigenvalue problem without any additional assumptions, and taking into consideration all possible wavenumbers. To reveal the feasibility of smooth stratified configuration in inclined two-phase flows, special attention was paid to multiple solution regions. The stability boundaries of each possible steady state solution (base flow) are presented on the flow pattern map accompanied by the wavenumbers and spatial profiles of the critical perturbations.

It is demonstrated that the long-wave stability analysis yields correct results for countercurrent liquid-liquid flows. However, this fact would not be established without the present analysis that considers all wavenumber perturbations. Although in air-water countercurrent flow short and intermediate wavelength perturbations are the most unstable ones, the long-wave boundary is located very close to the exact stability boundary for the lower solution, and due to low holdups and flow rates the exact analytical solution can be conveniently utilized with negligible inaccuracy. The present analysis confirms the feasibility of two stable solutions for the holdup in countercurrent liquid-liquid flows in a region of low flow rates. The region of stable lower holdup solution extends to higher upper layer flow rates, whereas upper solution is stable for higher lower layer flow rates. Countercurrent flows become unstable to the shear mode of instability, driven by shear in the dominating phase (forming the thick layer) at the wall and also at the interface. Although the perturbation patterns are similar along the stability boundaries of each of the solution, streamlines of the critically perturbed flow can have different configurations, owing to different base state velocity profiles. The waves with maximal amplitudes can be located either in the bulk of the thick layer or near the interface resulting in the generation of circulations cells. The stable regions for each of the holdup solutions shrink with increasing the inclination. Gradually, both flow configurations become unstable even for very low flow rates at inclination around 45° for the considered pair of liquids, which is in a good agreement with the experiments of Ullmann et al. (2003a). In gas-liquid (air-water) countercurrent flow only the lower holdup solution was found to be stable, and the stable region resembles that obtained in concurrent air-water down flow for the same channel inclination.

In slightly upward inclined flows the triple solution region is obtained for high light phase flow rates. Additional two stratified configurations with holdup lower than in single solution region (at lower gas flow rates) exist. The triple solutions are obtained when the gas flow rate is already sufficiently high to enable also a flow configuration where the entire liquid flow is dragged upward and backflow of the liquid near the bottom wall is avoided. Linear analysis reveals that the lower and middle holdup solutions are stable in the part of the triple solution region, while the upper solution is always unstable. The loss of stability in the flow with the low holdup is associated with the long-wavelength perturbations and occurs in the bulk of the air layer (similarly to horizontal flow). Moreover, the stability boundary for the lower holdup solution coincides with that of the corresponding horizontal flow, since the effect of inclination is negligible in this case. However, for the middle solution the short waves become more unstable, and the maximum disturbance amplitudes are observed both in the bulk of upper layer and at the interface. For the low holdup solution the streamlines of the critically perturbed flow are wavy in the air layer, and the highest amplitudes are in the vicinity of the interface leading to its deformation. For the middle holdup solution, a similar picture is observed for the streamlines in the upper phase and at the interface, while the lower phase is occupied by circulation cells owing to the backflow in the base flow. In both cases the critical perturbations arise in the flow propagate downstream.

In the case of downward flow, the triple solution region is obtained for high flow rates of the heavy phase, where none of the solutions are stable with respect to short-wave perturbations. In comparison to the horizontal flow, the downward inclination reduces drastically the stable region to very low liquid flow rates. The higher is the inclination angle, the smaller is the stable region. The most unstable perturbations are long waves with the patterns of shear instability in the gas phase. The streamlines for the critically perturbed flow are wavy in the air layer with maximal amplitudes either in its bulk, where circulation cells are generated, or near the deformed (wavy) interface. Similarly to upward flow, in both cases the critical perturbations propagate downstream.

The results of the present exact (linear) stability analysis can also be used to identify the necessary modifications that should be introduced in Two-Fluid models, which are still necessary in the more complicated pipe geometry, in order to improve the prediction of the stability boundary of smooth stratified flow via the simplified one dimensional approach.